\documentclass[pra, 10pt,aps, twocolumn,  notitlepage, showpacs, floatfix]{revtex4-1}
 
\pdfoutput=1
\usepackage{graphicx}
\usepackage{amsmath, amssymb, amsfonts, bm}
\usepackage{bbm}
\usepackage{latexsym}
\usepackage[colorlinks]{hyperref}
\usepackage{xcolor}
\graphicspath{{Figures/}}
\usepackage[final]{changes}

\newcommand{\D}[1]{\,\text{d}#1\,}

\newcommand{\fixme}[1]%
   {\begingroup{\color{blue}[NOTE: \textit{#1}]}\endgroup}

\begin{document}

\title{Photon blockade and the quantum-to-classical transition in the driven-dissipative Josephson pendulum coupled to a resonator}

\author{I.~Pietik\"ainen$^1$}
\author{J.~Tuorila$^{1,2}$}
\author{D.~S. Golubev$^2$}
\author{G.~S.~Paraoanu$^2$}

\affiliation{$^1$Nano and molecular systems research unit, University of Oulu, P.O. Box 3000, FI-90014 Oulu, Finland}
\affiliation{$^2$QTF Centre of Excellence, Department of Applied Physics, Aalto University, P.O. Box 15100, FI-00076 Aalto, Finland}

\date{\today}

\begin{abstract}
We investigate the driven quantum phase transition between \added{the} oscillating motion and the classical nearly free rotations of the Josephson pendulum coupled to a harmonic oscillator in the presence of dissipation. \added{We refer to this as the Josephson--Rabi model.} This model describes the standard setup of circuit quantum electrodynamics, where typically a transmon device is embedded in a superconducting cavity.
We find that by treating the system quantum mechanically this transition occurs at higher drive powers than expected from an all-classical treatment, which is a consequence of the quasiperiodicity originating in the discrete energy spectrum of the bound states. 
We calculate the photon number in the resonator and show that its dependence on the drive power is nonlinear. In addition, the resulting multi-photon blockade phenomenon is sensitive to the truncation of the number of states in the transmon, which limits the applicability of the standard Jaynes--Cummings model as an approximation for the pendulum-oscillator system. \added{We calculate the \textit{n}th order correlation functions of the blockaded microwave photons and observe the differences between the rotating-wave approximation and the full multilevel Josephson--Rabi Hamiltonian with the counter-rotating terms included. Finally, we} compare two different approaches to dissipation, namely the Floquet--Born--Markov and the Lindblad formalisms.  
\end{abstract}

\maketitle

\section{Introduction}
The pendulum, which can be seen as a rigid rotor in a gravitational potential~\cite{baker2005}, is a quintessential nonlinear system. It has two extreme dynamical regimes: the low-energy regime, where it can be approximated as a weakly anharmonic oscillator, and the high-energy regime, where it behaves as a free rotor. Most notably, the pendulum physics appears in systems governed by the Josephson effect, where the Josephson energy is analogous to the gravitational energy, and the role of the momentum is taken by the imbalance in the number of particles due to tunneling across the weak link. Such a system is typically referred to as a Josephson pendulum. 

In ultracold degenerate atomic gases, several realizations of the Josephson pendulum have been studied~\cite{Leggett2001,paraoanu2001, Smerzi1997, Marino1999}. While the superfluid-fermion case~\cite{Paraoanu2002,Heikkinen2010} still awaits experimental realization, the bosonic-gas version has been already demonstrated~\cite{Albiez2005, Levy2007}. Also in this case two regimes have been identified: small Josephson oscillations, corresponding to the low-energy limit case described here, and the macroscopic self-trapping regime~\cite{Smerzi1997, Marino1999}, corresponding to the free-rotor situation. 
Another example is an oscillating $LC$ electrical circuit \added{with a nonlinear inductance realized} as a tunnel barrier between two superconducting leads.  This is the case of the transmon circuit~\cite{koch2007}, which \deleted{will be the focus of this paper. The transmon }is currently one of the most promising approaches to quantum processing of information, with high-fidelity operations and good prospects for scalability. 
Its two lowest eigenstates are close to those of a harmonic oscillator, with only weak perturbations caused by the anharmonicity of the potential. 
The weak anharmonicity also guarantees that the lowest states of the transmon are immune to charge noise, which is a major source of decoherence in superconducting quantum circuits. 

In this paper we consider a paradigmatic model which arises when the Josephson pendulum is interacting with a resonator. Circuit quantum electrodynamics offers a rigorous embodyment of the above model as a transmon device 
coupled to a superconducting resonator - fabricated either as a three-dimensional cavity or as a coplanar waveguide segment. In this realization, the system is driven by an external field of variable frequency, and dissipation affects both the transmon and the resonator. 

We study in detail the onset of nonlinearity in the driven-dissipative phase transition between the quantum and the classical regimes. \deleted{We concentrate especially on the low-energy regime of the transmon-resonator system, which has received only a little attention in the literature.}
We further compare the photon number in detail with the corresponding transmon occupation, and demonstrate that the onset of nonlinearities is accompanied by the excitation of all bound states of the transmon and, thus, it is sensitive to the transmon truncation. 
We also find that the onset of the nonlinearities is sensitive to the energy level structure of the transmon, {\it e.g.} on the gate charge which affects the eigenenergies near and outside the edge of the transmon potential. 
The results also show that the full classical treatment is justified only in the high-amplitude regime, yielding significant discrepancies in the low-amplitude regime -  where the phenomenology is governed by photon blockade. This means that the system undergoes a genuine quantum-to-classical phase transition. \added{Calculations of the \textit{n}th order correlation functions support this conclusion. These correlations are very small in the blockaded regime and increase beyond unity in the classical one. Moreover, the effects of the counter-rotating (Bloch--Siegert) terms in the full multilevel Josephson--Rabi system are clearly observable.
Overall,} our numerical simulations demonstrate that the multi-photon blockade phenomenon is qualitatively different for a realistic multilevel anharmonic system compared to the Jaynes--Cummings case studied extensively in the literature.  

\added{To introduce dissipation we use two models,} namely the conventional Lindblad master equation and the Floquet--Born--Markov master equation, which is developed especially to capture the effects of the drive on the dissipation. We show that both yield relatively  close results. However, we emphasize that the Floquet--Born--Markov approach should be preferred because its numerical implementation is considerably more efficient than that of the corresponding Lindblad equation.

While our motivation is to elucidate fundamental physics, 
in the burgeoning field of quantum technologies several applications of our results can be envisioned. For example, the single-photon blockade can be employed to realize single-photon sources, and the two-photon blockade can be utilized to produce transistors controlled by a single photon and filters that yield correlated two-photon outputs from an input of random photons~\cite{Kubanek2008}. In the field of quantum simulations, the Jaynes--Cummings model can be mapped \added{onto} the Dirac electron in an electromagnetic field, with the coupling and the drive amplitude corresponding respectively to the magnetic and electric field: \added{then,} the photon blockade regime is associated with a discrete spectrum of the Dirac equation, while the breakdown of the blockade corresponds to a continuous spectrum~\cite{Gutierrez-Jauregui2018}. Finally, the switching behavior of the pendulum in the transition region can be used for designing bifurcation amplifiers for the single-shot nondissipative readout of qubits~\cite{Vijay2009}.

The paper is organized as follows. In Section~\ref{sec:II} we introduce the electrical circuit which realizes the pendulum-oscillator system, 
calculate its eigenenergies, 
identify the two dynamical regimes of the small oscillations and the free rotor. 
In Section~\ref{sec:III}, we introduce the drive and dissipation. We discuss two formalisms for dissipation, namely the Lindblad equation and the Floquet--Born--Markov approach. Section~\ref{sec:IV} presents the main results for the quantum-to-classical transition and the photon blockade, focusing on the resonant case. Here, we also discuss the gate dependence, the ultra-strong coupling regime, \added{and we obtain the \textit{n}th order photon correlations.} Section~\ref{sec:V} is dedicated to conclusions. 

\section{Circuit-QED implementation of a Josephson pendulum coupled to a resonator}\label{sec:II}

\begin{figure}[h!]
\includegraphics[width=1.0\linewidth]{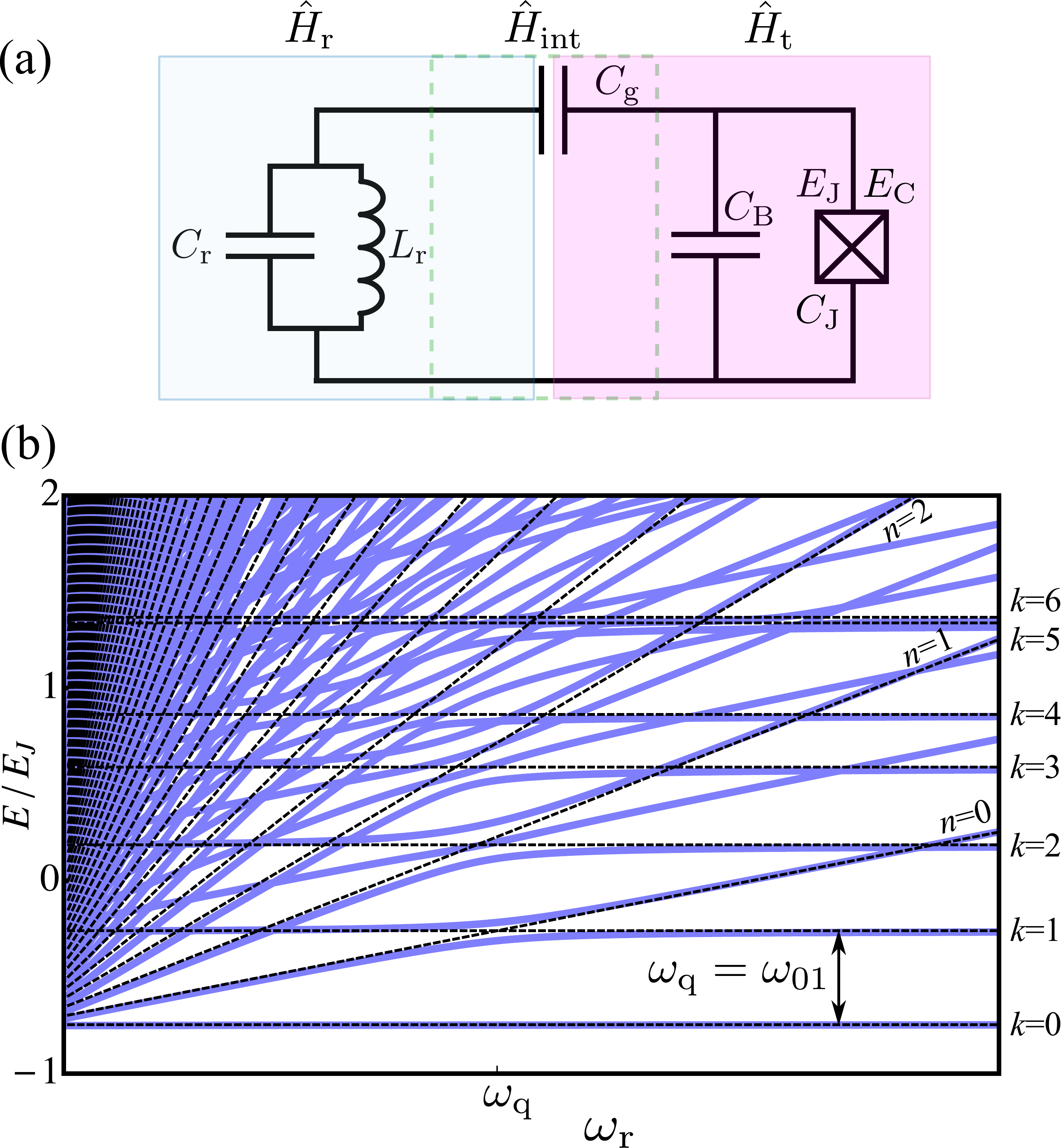}\hfill
\caption{Electrical circuit and the corresponding eigenenergy spectrum. (a) Lumped-element schematic of a transmon-resonator superconducting circuit. The resonator and the transmon are marked with blue \added{(left)} and magenta \added{(right)} rectangles. 
(b) Numerically obtained eigenenergies of the resonator-pendulum Hamiltonian in Eq.~(\ref{eq:H0}) are shown in blue \added{(gray)} as a function of the resonator frequency. The bare pendulum eigenenergies $\hbar \omega_k$ are denoted with dashed horizontal lines and indicated with the label $k$. The eigenenergies of the uncoupled system, defined \replaced{in}{by} Eq.~(\ref{eq:Hr}) and Eq.~(\ref{eq:transmonHam}), are given by the dashed lines whose slope increases in integer steps with the number of quanta in the oscillator as $n\hbar \omega_{\rm r}$. We only show the eigenenergies of the uncoupled system for the case of pendulum in its ground state, but we note that one obtains a similar infinite fan of energies for each pendulum eigenstate.
Note that in general $\omega_{\rm q}\neq \omega_{\rm p}$. We have used the parameters in Table~\ref{tab:params1} 
and fixed $n_{\rm g}=0$. 
}
\label{fig:oscpendevals}
\end{figure}

We discuss here the physical realization of the Josephson pendulum-resonator system as an electrical circuit consisting of a transmon device coupled capacitively to an $LC$ oscillator, as depicted in Fig.~\ref{fig:oscpendevals}(a). 
The coupled system is modeled by the Hamiltonian
\begin{equation}\label{eq:H0}
\hat H_{0} = \hat H_{\rm r} + \hat H_{\rm t} + \hat H_{\rm c},
\end{equation}
where 
\begin{eqnarray}
\hat H_{\rm r} &=& \hbar\omega_{\rm r}\hat a^{\dag}\hat a, \label{eq:Hr}\\
\hat H_{\rm t} &=& 4E_{\rm C}(\hat n-n_{\rm g})^2 - E_{\rm J}\cos \hat \varphi, \label{eq:transmonHam}\\
\hat H_{\rm c} &=& \hbar g\hat n(\hat a^{\dag}+\hat a)
\end{eqnarray}
describe the resonator, the transmon, and their coupling, respectively.
We have defined $\hat a$ as the annihilation operator of the harmonic oscillator, and used $\hat n = -i\partial /\partial\varphi$ as the conjugate momentum operator of the superconducting phase difference $\hat \varphi$. 
These operators obey the canonical commutation relation $[\hat \varphi,\hat n]=i$. The angular frequency of the resonator 
is given by $\omega_{\rm r}$. 
We have also denoted the Josephson energy with $E_{\rm J}$, and the charging energy with $E_{\rm C } = e^2/(2C_\Sigma)$ where the capacitance on the superconducting island of the transmon is given as $C_{\Sigma}=C_{\rm B} + C_{\rm J} + C_{\rm g}$. Using the circuit diagram in Fig.~\ref{fig:oscpendevals}(a), we obtain the coupling constant $g = 2 e C_{\rm g}q_{\rm zp}/(\hbar C_{\Sigma} C_{\rm r})$, where the zero-point fluctuation amplitude of the oscillator charge is denoted with $q_{\rm zp} = \sqrt{C_{\rm r}\hbar \omega_{\rm r}/2}$~\cite{koch2007}. 

Let us briefly discuss the two components of this system: the Josephson pendulum and the resonator. The pendulum physics is realized by the superconducting transmon circuit~\cite{koch2007} in Fig.~\ref{fig:oscpendevals}(a) and described by the Hamiltonian $\hat H_{\rm t}$ in Eq.~(\ref{eq:transmonHam}).
As discussed in Ref.~\cite{koch2007}, the Hamiltonian of the transmon is analogous to that of an electrically charged particle whose motion is restricted to a circular orbit and subjected to homogeneous and perpendicular gravitational and magnetic fields.
By fixing the x and z directions as those of the gravity and the magnetic field, respectively, the position of the particle is completely determined by the motion along the polar angle in the xy plane. The polar angle can be identified as the $\varphi$ coordinate of the pendulum.
Thus, the kinetic energy part of the Hamiltonian~(\ref{eq:transmonHam}) describes a free rotor in a homogeneous magnetic field.
In the symmetric gauge, the vector potential of the field imposes an effective constant shift 
for the $\varphi$ component of the momentum, which is analogous to the offset charge $n_{\textrm g}$ on the superconducting island induced either by the environment or by a gate voltage.
In the following, the 'plasma' frequency for the transmon is given by $\omega_{\rm p} = \sqrt{8E_{\rm C}E_{\rm J}}/\hbar$ and describes the classical oscillations of the linearized circuit. The parameter $\eta=E_{\rm J}/E_{\rm C}$ is the ratio between the potential and kinetic energy of the pendulum, and determines, by the condition $\eta \gg 1$, whether the device is in the charge-insensitive regime
of the free rotor and the gravitational potential.  

The eigenvalues $\{\hbar \omega_k\}$ and the corresponding eigenvectors $\{|k\rangle\}$, with $k=0,1,\ldots$, of the Hamiltonian in Eq.~(\ref{eq:transmonHam}) can be obtained by solving the Mathieu equation, see Appendix~\ref{app:eigenvalue}.
In general, the eigenvalues of the coupled system Hamiltonian $\hat H_0$ in Eq.~(\ref{eq:H0}) have to be solved numerically. With a sufficient truncation in the Hilbert spaces of the uncoupled systems in Eq.~(\ref{eq:Hr}) and Eq.~(\ref{eq:transmonHam}), one can represent the Hamiltonian $\hat H_0$ in a matrix form. The resulting 
eigenvalues of the truncated Hamiltonian $\hat H_0$ are shown in Fig.~\ref{fig:oscpendevals}. We see that the coupling creates avoided crossings at the locations where the pendulum transition frequencies are equal to positive integer multiples of the resonator quantum. Also, the density of states increases drastically with the energy. The nonlinearity in the system is characterized by the non-equidistant spacings between the energy levels. Their origin is the sinusoidal Josephson potential of the transmon. 
Here, we are especially interested in the regime where the resonator frequency is (nearly) resonant with the frequency of the lowest transition of the pendulum, i.e. when $\omega_{\rm r}\approx\omega_{\rm q} = \omega_{01}=\omega_1-\omega_0$.

The Hamiltonian $\hat H_0$ in Eq.~(\ref{eq:H0}) can be represented in the eigenbasis $\{|k\rangle\}$ of the Josephson pendulum as
\begin{equation}
	\hat{H}_{0} = \hbar\omega_{\rm r} \hat{a}^{\dag}\hat{a} + \sum_{k=0}^{K-1} \hbar\omega_{k} \vert k\rangle\langle k\vert + \hbar g(\hat{a}^{\dag}+\hat{a})\sum_{k,\ell=0}^{K-1}\hat{\Pi}_{k\ell}.\label{eq:ManyStatesHam}
\end{equation}
\added{We refer to this as the Josephson--Rabi Hamiltonian.} Here, $K$ is the number of transmon states included in the truncation. We have also defined 
$\hat\Pi_{k\ell}\equiv \langle k|\hat{n}|\ell\rangle |k\rangle\langle \ell|$ which is the representation of the Cooper-pair-number operator in the eigenbasis of the transmon. 

A useful classification of the eigenstates can be obtained by using the fact that the transmon can be approximated as a weakly anharmonic oscillator \cite{koch2007}, thus $\langle k|\hat{n}|\ell\rangle$ is negligible if $k$ and $\ell$ differ by more than 1. Together with the rotating-wave approximation, this results in
\begin{equation}
\begin{split}
\hat{H}_{0} \approx &\hbar\omega_{\rm r} \hat{a}^{\dag}\hat{a} + \sum_{k=0}^{K-1} \hbar\omega_{k}\vert k\rangle\langle k\vert \\
&+ \hbar g\sum_{k=0}^{K-2} \left(\hat{a}\hat{\Pi}_{k,k+1}^{\dag} +  \hat{a}^{\dag} \hat{\Pi}_{k,k+1} \right),\label{eq:ManyStatesHam_simple}
\end{split}
\end{equation}
Here, we introduce the total excitation-number operator as
\begin{equation}
	\hat N = \hat a^{\rm \dag}\hat a + \sum_{k=0}^{K-1} k\vert k\rangle\langle k\vert,\label{eq:exitationN_K}
\end{equation}
which commutes with the Hamiltonian in Eq.~(\ref{eq:ManyStatesHam_simple}). Thus, the eigenstates of this Hamiltonian can be labeled by the eigevalues of $\hat N$, which is a representation that we will find useful \replaced{in the discussion of the}{later when discussing the} transitions between these states. The terms neglected in the rotating-wave approximation can be treated as small perturbations except for transitions where the coupling frequency $g_{\ell k} = g\langle k|\hat n|\ell\rangle$  becomes a considerable fraction of the corresponding transition frequency $\omega_{\ell k}=\omega_{k}-\omega_\ell$ and, thus, enters the ultrastrong coupling regime with $g_{k\ell} \geq 0.1\times \omega_{\ell k}$. In the ultrastrong coupling regime and beyond, the eigenstates are superpositions of states with different excitation numbers and cannot, thus, anymore be labeled with $N$.

Another important approximation for the Hamiltonian in Eq.(\ref{eq:ManyStatesHam}) is the two-state truncation ($K=2$), which reduces it to the Rabi Hamiltonian
\begin{equation}\label{eq:HRabi}
	\hat H_{\rm R} = \hbar \omega_{\rm r}\hat a^{\dag}\hat a+\hbar \omega_{\rm q} \hat \sigma_+\hat \sigma_- + \hbar g_{01}(\hat a^{\dag}+\hat a)\hat \sigma_{\rm x}.
\end{equation}
Here $g_{01}=g\langle 1|\hat n|0\rangle$, the qubit annihilation operator is $\hat \sigma_- = |0\rangle\langle 1|$,  and the Pauli spin matrix $\hat \sigma_{\rm x}=\hat \sigma_-+\hat \sigma_+$.  
The Rabi Hamiltonian is a good approximation to the pendulum-oscillator system as long as the corrections for the low-energy eigenvalues and eigenstates, arising from the higher excited states of the pendulum, are taken properly into account~\cite{boissonneault2009a,boissonneault2012b,boissonneault2012c}.

Further, by performing a rotating-wave approximation, we obtain the standard Jaynes--Cummings model
\begin{equation}
	\hat H_{\rm JC} = \hbar \omega_{\rm r}\hat a^{\dag}\hat a+\hbar \omega_{\rm q} \hat \sigma_+\hat \sigma_- + \hbar g_{01}(\hat a^{\dag}\hat \sigma_-+\hat a\hat \sigma_+),\label{eq:HJC}
\end{equation}
which also results from a truncation of Eq.~(\ref{eq:ManyStatesHam_simple}) to the low-energy subspace spanned by the lowest two eigenstates of the transmon.
Apart from the non-degenerate ground state $|0,0\rangle$ with zero energy, the excited-state eigenenergies of the Jaynes--Cummings Hamiltonian in Eq.~(\ref{eq:HJC}) form a characteristic doublet structure. In the resonant case, the excited-state eigenenergies and the corresponding eigenstates are given by
\begin{eqnarray}
	E_{n_{r},\pm} &=& n_{r}\hbar \omega_{\rm r} \pm \sqrt{n_{r}}\hbar g_{01}, \label{eq:JCener}\\
	|n_{r},\pm\rangle &=& \frac{1}{\sqrt{2}}(|n_{r},0\rangle \pm |n_{r}-1,1\rangle). \label{eq:JCstates}
\end{eqnarray}
Here, $n_r=1,2,\ldots$ and we have denoted eigenstates of the uncoupled Jaynes--Cummings Hamiltonian with $\{|n_{r},0\rangle , |n_{r},1\rangle\}$
where $|n_{r}\rangle$ are the eigenstates of the resonator with $n_{r}=0,1,\ldots$. 

Due to the rotating-wave approximation, the Jaynes--Cummings Hamiltonian commutes with the excitation-number operator in Eq.~(\ref{eq:exitationN_K}) truncated to two states and represented as
\begin{equation}
	\hat N = \hat a^{\rm \dag}\hat a + \hat \sigma_+\hat \sigma_-. \label{eq:exitationN_2}
\end{equation}
Thus, they have joint eigenstates and, in addition, the excitation number $N$ is a conserved quantity. For a doublet with given $n_{r}$, the eigenvalue of the excitation-number operator is $N=n_{r}$, while for the ground state $N=0$. We note that the transition energies between the Jaynes--Cummings eigenstates depend nonlinearly on $N$. Especially, the transition energies from the ground state $\vert 0,0\rangle$ to the eigenstate $\vert n_{r},\pm \rangle$ are given by $n_{r}\hbar\omega_{\rm r} \pm \sqrt{n_{r}}\hbar g_{01}$.

\section{Models for the driven-dissipative Josephson pendulum coupled to the harmonic oscillator}\label{sec:III}

Here, we provide a master equation approach that incorporates the effects of the drive and dissipation to the coupled system.
Previous studies on this system have typically truncated the transmon to the low-energy subspace spanned by the two lowest energy eigenstates~\cite{Bishop2010, Reed2010}, or treated the dissipation in the conventional Lindblad formalism~\cite{bishop2009}. Recent studies~\cite{pietikainen2017, pietikainen2018, verney2018, lescanne2018} have treated the dissipation at the detuned limit using the Floquet--Born--Markov approach. We will apply a similar formalism for the case where the pendulum and resonator are in resonance in the low-energy subspace. Especially, we study the driven-dissipative transition between the low-energy and the free rotor regimes of the pendulum in terms of the dependence of 
the number $N_{\rm r}$ of quanta in the resonator on the drive power.

\subsection{Coupling to the drive}

The system shown in Fig.~\ref{fig:oscpendevals} and described by the Hamiltonian in Eq.~(\ref{eq:H0}) can be excited by coupling the resonator to a monochromatic driving signal modeled with the Hamiltonian
\begin{equation}\label{eq:Hd}
	\hat H_{\rm d} = \hbar A \cos(\omega_{\rm d}t)[\hat a^{\dag}+\hat a],
\end{equation}
where $A$ and $\omega_{\rm d}$ are the amplitude and the angular frequency of the drive, respectively. This results in a total system Hamiltonian
$\hat H_{\rm S} = \hat H_{0}  + \hat{H}_{\rm d}$.
For low-amplitude drive, only the first two states of the pendulum have a significant occupation 
and, thus, the Hamiltonian $\hat H_0$ can be truncated into the form of the well-known Rabi Hamiltonian in Eq.~(\ref{eq:HRabi}), which in turn, under the rotating-wave approximation, yields the standard Jaynes--Cummings Hamiltonian in Eq.~(\ref{eq:HJC}).  
The transitions induced by the drive in the Jaynes--Cummings system are subjected to a selection rule -- the occupation number can change only by one, i.e. $N \rightarrow N\pm 1$. This follows from the relations
\begin{eqnarray}
\langle n_{r},\pm|(\hat a^{\dag}+\hat a)|0,0\rangle &=& \frac{1}{\sqrt{2}}\delta_{n_{r},1}, \label{eq:selection1}\\
\langle n_{r},\pm|(\hat a^{\dag}+\hat a)|\ell_{r},\pm \rangle &=& \frac{1}{2}\left(\sqrt{n_{r}}+\sqrt{n_{r}-1}\right)\delta_{n_{r},\ell_{r}+1} \nonumber\\
&+&\frac{1}{2}\left(\sqrt{n_{r}+1}+\sqrt{n_{r}}\right)\delta_{n_{r},\ell_{r}-1}\label{eq:selection2}
\end{eqnarray}
As a consequence, the system climbs up the Jaynes--Cummings ladder by one step at a time. Particularly, a system in the ground state is coupled directly only to states $|1,\pm\rangle$. Indeed, in such a system the Jaynes--Cummings ladder has been observed~\cite{fink2008}, as well as the effect of strong drive in the off-resonant~\cite{pietikainen2017} and on-resonant case~\cite{fink2017}. The Jaynes--Cummings model offers a good starting point for understanding the phenomenon of photon blockade in the pendulum-resonator system, which will be discussed later in detail. Indeed, it is apparent from Eq.~(\ref{eq:JCener}) that, as the system is driven externally by not too intense fields, the excitation to higher levels in the resonator is suppressed by the higher levels being off-resonant, due to the nonlinearity induced by the coupling. This is referred to as photon blockade. As the drive amplitude increases further, the entire Jaynes--Cummings hierarchy breaks down~\cite{carmichael2015}.

However, in weakly anharmonic systems such as the transmon, as the drive amplitude is increased, the higher excited states of the Josephson pendulum become occupied and the two-state approximation becomes insufficient. As a consequence, the system has to be modeled by a
\added{multilevel Josephson--Rabi model~\cite{pietikainen2017,pietikainen2018,lescanne2018,verney2018}, see Eq.~(\ref{eq:ManyStatesHam}), which includes the higher excitations of the transmon as well as the counter-rotating (Bloch--Siegert) terms.} In the resonant case, the need to take into account the second excited state of the transmon has been pointed out already in Ref.~\cite{fink2017}. \replaced{Here}{Next}, at larger drive amplitudes, the pendulum escapes the low-energy subspace defined by the states localized in a well of the cosine potential and the unbound free rotor states also become occupied~\cite{pietikainen2017,pietikainen2018,lescanne2018,verney2018} even in the case of strongly detuned drive frequency.  In the limit of very high drive power, the pendulum behaves as a free rotor and the nonlinear potential can be neglected.
Consequently, the resonance frequency of the system is set by the bare resonator frequency, instead of the normal modes.

\subsection{Dissipative coupling}

The dissipation is treated by modeling the environment \added{as a} thermal bosonic bath which is coupled bilinearly to the resonator. 
The Hamiltonian of the driven system coupled to the bath can be written as
\begin{equation}\label{eq:totHam}
\hat H = \hat H_{\rm S} + \hat H_{\rm B} + \hat H_{\rm int},
\end{equation}
where
\begin{eqnarray}
\hat H_{\rm B} &=& \hbar \sum_k \Omega_k\hat b_k^{\dag}\hat b_k,\\
\hat H_{\rm int} &=& \hbar (\hat a^{\dag}+\hat a) \sum_k g_k(\hat b_k^{\dag}+\hat b_k).\label{eq:dissint}
\end{eqnarray}
Above, $\{\hat b_k\}$, $\{\Omega_k\}$, and $\{g_k\}$ are the annihilation operators, the angular frequencies, and the coupling frequencies of the bath oscillators. 
We use this model in the derivation of a master equation for the reduced density operator of the system. 
We proceed in the conventional way and assume the factorized initial state $\hat \rho(0) = \hat\rho_{\rm S}(0)\otimes \hat \rho_{\rm B}(0)$, apply the standard Born and Markov approximations, trace over the bath, and perform the secular approximation. 
As a result, we obtain a master equation in the standard Lindblad form.

\subsection{Lindblad master equation}\label{sec:Lindblad}

Conventionally, the dissipation in the circuit QED setup has been treated using independent Lindblad dissipators for the resonator and for the pendulum. Formally, this can be achieved by coupling the pendulum to another heat bath formed by an infinite set of harmonic oscillators. This interaction can be described with the Hamiltonian
\begin{equation}\label{eq:transdissint}
\hat H_{\rm int}^{\rm t} = \hbar \hat n \sum_k f_k(\hat c_k^{\dag}+\hat c_k),
\end{equation}
where $\{f_k\}$ and $\{\hat c_k\}$ are the coupling frequencies and the annihilation operators of the bath oscillators. The bath is coupled to the transmon through the charge operator $\hat n$ which is the typical source of decoherence in the charge-based superconducting qubit realizations.
By following the typical Born--Markov derivation of the master equation for the uncoupled subsystems, one obtains a Lindblad equation where the dissipators induce transitions between the eigenstates of the uncoupled ($g=0$) system~\cite{Breuer2002, scala2007, beaudoin2011, tuorila2017} 
\begin{equation}
\begin{split}
\frac{\D {\hat\rho}}{\D t} =& -\frac{i}{\hbar}[\hat{H}_{\rm S},\hat{\rho}] +\kappa[n_{\rm th}(\omega_{\rm r})+1]\mathcal{L}[\hat{a}]\hat\rho \\
&+\kappa n_{\rm th}(\omega_{\rm r})\mathcal{L}[\hat{a}^\dagger]\hat\rho \\
&+\sum_{k\ell} \Gamma_{k\ell}\mathcal{L}[|\ell\rangle\langle k|]\hat{\rho},
\end{split} 
\label{eq:LindbladME}
\end{equation}
where  $\mathcal{L}[\hat{A}]\hat\rho = \frac12 (2\hat{A}\hat\rho \hat{A}^{\dag} - \hat{A}^{\dag}\hat{A}\hat\rho-\hat\rho \hat{A}^{\dag}\hat{A})$ is the Lindblad superoperator and $n_{\rm th}(\omega)=1/[e^{\hbar\omega/(k_{\rm B} T)}-1]$ is the Bose--Einstein occupation. Note that the treatment of dissipation as superoperators acting separately on the qubit and on the resonator is valid if their coupling strength and the drive amplitude are weak compared to the transition frequencies of the uncoupled system-environment. 
Above, we have also assumed an ohmic spectrum for the resonator bath. 

In the Lindblad master equation~(\ref{eq:LindbladME}), we have included the effects arising from the coupling $g$ and the drive into the coherent von Neumann part of the dynamics. The first two incoherent terms cause transitions between the eigenstates of the resonator and arise from the interaction Hamiltonian in Eq.~(\ref{eq:dissint}). The strength of this interaction is characterized with the spontaneous emission rate $\kappa$.  The last term describes the relaxation, excitation, and dephasing of the transmon caused by the interaction Hamiltonian in Eq.~(\ref{eq:transdissint}). The transition rates $\Gamma_{k\ell}$ between the transmon eigenstates follow the Fermi's golden rule as
\begin{equation}
\Gamma_{k\ell} = |\langle \ell | \hat n| k\rangle|^2 S(\omega_{k\ell}).
\end{equation}
In our numerical implementation, we have assumed that the fluctuations of the transmon bath can also be characterised with an ohmic spectrum $S(\omega)=\frac{\gamma_0\omega}{1-\exp[-\hbar\omega/k_{\rm B}T]}$, where $\gamma_0$ is a dimensionless factor describing the bath-coupling strength. We have also denoted the transition frequencies of the transmon with $\omega_{k\ell} = \omega_{\ell}-\omega_k$. 

Here, the magnitude of the transition rate from state $|k\rangle$ to the state $|\ell\rangle$ is given by the corresponding matrix element of the coupling operator $\hat n$ and the coupling strength $\gamma_0$. 
We note that in a typical superconducting resonator-transmon realization one has $\gamma=\gamma_0 \omega_{01}\ll \kappa$. In this so-called bad-cavity limit, the effects of the transmon bath are negligible especially if the coupling frequency $g$ with the resonator is large. Thus, the main contribution of the transmon dissipators in the master equation Eq.~(\ref{eq:LindbladME}) is that it results to faster convergence in the numerical implementation of the dynamics.


\subsection{Floquet--Born--Markov formalism}\label{sec:FBM}

The dissipators in the Lindblad model above are derived under the assumption of weak driving and weak coupling between the transmon and the resonator. 
However, both the driving and the coupling affect the eigenstates of the system and, thus, 
have to be taken into account in the derivation of the master equation. 
This can be achieved in the so-called Floquet--Born--Markov approach, where the drive and the transmon-resonator coupling are explicitly included throughout the derivation of the dissipators~\cite{tuorila2013, pietikainen2017,pietikainen2018,lescanne2018,verney2018}. For this purpose, we represent the system in terms of the quasienergy states which can be obtained only numerically. 

Since the drive in Eq.~(\ref{eq:Hd}) is $\tau=2\pi/\omega_{\rm d}$-periodic, the solution to the time-dependent Schr\"odinger equation
\begin{equation}\label{eq:tdse}
i\hbar\frac{\D}{\D t}|\Psi(t)\rangle = \hat{H}_{\rm S}(t) |\Psi(t)\rangle,
\end{equation} 
corresponding to the Hamiltonian $\hat{H}_{\rm S}(t)$ in Eq.~(\ref{eq:totHam}), can be written in the form
\begin{equation}\label{eq:FloqState}
|\Psi(t)\rangle = e^{-i\varepsilon t/\hbar} |\Phi(t)\rangle,
\end{equation}
where $\varepsilon$ are the quasienergies and $|\Phi(t)\rangle$ are the corresponding $\tau$-periodic quasienergy states.
By defining the unitary time-propagator as
\begin{equation}\label{eq:FloqProp}
\hat{U}(t_2,t_1)|\Psi(t_1)\rangle =|\Psi(t_2)\rangle,
\end{equation}
one can rewrite the Schr\"odinger equation~(\ref{eq:tdse}) in the form
\begin{equation}
i\hbar\frac{\D}{\D t}\hat{U}(t,0) = \hat{H}_{\rm S}(t)\hat{U}(t,0).
\end{equation}
Using Eqs.~(\ref{eq:FloqState}) and (\ref{eq:FloqProp}), we obtain
\begin{eqnarray}
\hat{U}(\tau,0)|\Phi(0)\rangle &=& e^{-i\varepsilon \tau/\hbar} |\Phi(0)\rangle, \label{eq:QEproblem}
\end{eqnarray}
from which the quasienergies $\varepsilon_\alpha$ and the corresponding quasienergy states $|\Phi_\alpha(0)\rangle$ can be solved. Using the propagator $\hat U$, one can obtain the quasienergy states for all times from
\begin{equation}
\hat{U}(t,0)|\Phi_\alpha(0)\rangle = e^{-i\varepsilon_\alpha t/\hbar} |\Phi_\alpha(t)\rangle.
\end{equation}
Due to the periodicity of $|\Phi_\alpha(t)\rangle$, it is sufficient to find the quasienergy states for the time interval $t\in[0,\tau ]$. Also, if $\varepsilon_\alpha$ is a solution for Eq.~(\ref{eq:QEproblem}), then $\varepsilon_\alpha +\ell\hbar\omega_{\rm d}$ is also a solution. Indeed, all  solutions of Eq.~(\ref{eq:QEproblem}) can be obtained from the solutions of a single energy interval of $\hbar\omega_{\rm d}$. These energy intervals are called Brillouin zones, 
in analogy with the terminology used in solid-state physics for periodic potentials.

The master equation for the density operator in the quasienergy basis can be written as~\cite{Blumel1991, Grifoni1998}
\begin{equation}\label{eq:FBM}
\begin{split}
\dot{\rho}_{\alpha\alpha}(t) &= \sum_{\nu} \left[\Gamma_{\nu\alpha}\rho_{\nu\nu}(t)-\Gamma_{\alpha\nu}\rho_{\alpha\alpha}(t)\right],\\
\dot{\rho}_{\alpha\beta}(t) &= -\frac12 \sum_{\nu}\left[\Gamma_{\alpha\nu}+\Gamma_{\beta\nu}\right]\rho_{\alpha\beta}(t), \ \ \alpha\neq \beta,
\end{split}
\end{equation}
where
\begin{equation}
\begin{split}
\Gamma_{\alpha\beta}&=\sum_{\ell=-\infty}^{\infty} \left[\gamma_{\alpha\beta \ell}+n_{\rm th}(|\Delta_{\alpha\beta \ell}|)\left(\gamma_{\alpha\beta \ell}+\gamma_{\beta \alpha -\ell}\right)\right],\\
\gamma_{\alpha\beta \ell} &= \frac{\pi}{2} \kappa \theta(\Delta_{\alpha\beta\ell})\frac{\Delta_{\alpha\beta\ell}}{\omega_{\rm r}}|X_{\alpha\beta\ell}|^2.
\end{split}
\end{equation}
Above, $\theta(\omega)$ is the Heaviside step-function and $\hbar\Delta_{\alpha \beta \ell} = \varepsilon_{\alpha} - \varepsilon_{\beta} + \ell\hbar\omega_{\rm d}$
is the energy difference between the states $\alpha$ and $\beta$ in Brillouin zones separated by $\ell$. Also, 
\begin{equation}
X_{\alpha\beta \ell} = \frac{1}{\tau}\int_{t_0}^{t_0 +\tau} \D t e^{-i\ell\omega_d t} \langle \Phi_\alpha(t)|(\hat{a}^\dagger+\hat{a})|\Phi_\beta(t)\rangle,
\end{equation}
where $t_0$ is some initial time after the system has reached a steady state.

From Eq.~(\ref{eq:FBM}), we obtain the occupation probabilities $p_\alpha=\rho_{\alpha\alpha}(t\rightarrow \infty)$ in the steady state as 
\begin{equation}
p_{\alpha} = \frac{\sum_{\nu\neq \alpha} \Gamma_{\nu\alpha}p_{\nu}}{\sum_{\nu\neq \alpha}\Gamma_{\alpha\nu}},
\end{equation}
and the photon number
\begin{equation}\label{eq:FBMNr}
N_{\rm r} = \sum_\alpha p_\alpha\langle \hat{a}^\dagger\hat{a}\rangle_\alpha,
\end{equation}
where 
\begin{equation}
\langle \hat{a}^\dagger\hat{a}\rangle_\alpha= \frac{1}{\tau}\int_{t_0}^{t_0 +\tau} \D t  \langle \Phi_\alpha(t)|\hat{a}^\dagger\hat{a}|\Phi_\alpha(t)\rangle,
\end{equation}
is the photon number in a single quasienergy state. The occupation probability for the transmon state $\vert k\rangle$ is given by
\begin{equation}\label{eq:FBMPk}
P_k= \frac{1}{\tau}\sum_\alpha p_\alpha\int_{t_0}^{t_0 +\tau} \D t  \langle \Phi_\alpha(t)|k\rangle\langle k|\Phi_\alpha(t)\rangle \,.
\end{equation}
We emphasize that this method assumes weak coupling to the bath but no such restrictions are made for the drive and pendulum-resonator coupling strengths. 
As a consequence, the dissipators induce transitions between the quasienergy states of the driven coupled system.


\subsection{Parameters}

The parameter space is spanned by seven independent parameters which are shown in Table~\ref{tab:params1}.
\begin{table}[th]
\begin{tabular}{|ccc|}\hline
Symbol & Parameter  & Value\\ \hline
$\omega_{\rm q}$ & qubit frequency & 1.0\\
$\omega_{\rm d}$ & drive frequency & 0.98\\
$\omega_{\rm p}$ & plasma oscillation frequency & 1.08\\
$g$ & coupling frequency & 0.04\\
$\kappa$ & resonator dissipation rate & 0.002\\
$k_{\rm B}T$ & thermal energy & 0.13\\
$E_{\rm C}$ & charging energy & 0.07\\
$\eta$  & energy ratio $E_{\rm J}/E_{\rm C}$ & 30 \\
\hline
\end{tabular}
\caption{Parameters of the driven and dissipative oscillator-pendulum system. The numerical values of the angular frequencies and energies used in the numerical simulations are given in units of $\omega_{\rm r}$ and $\hbar\omega_{\rm r}$, respectively. We note that $\omega_{\rm q}$ is determined by $E_{\rm C}$ and $\eta$, see the text.}
\label{tab:params1}
\end{table}
We fix the values of the energy ratio $\eta=E_{\rm J}/E_{\rm C}$ and the coupling strengths $g$ and $\kappa$. The ratio $\eta$ sets the number $K_{\rm b}$ of bound states in the pendulum, see Appendix \ref{app:eigenvalue}, but does not qualitatively affect the response. We have used a moderate value of $\eta$ in the transmon regime, in order to keep $K_{\rm b}$ low allowing more elaborate discussion of the transient effects between the low-energy oscillator and rotor limits. We use the Born, Markov, and secular approximations in the description of dissipation which means that the value of $\kappa$ has to be smaller than the system frequencies. In addition, we work in the experimentally relevant strong coupling regime where the oscillator-pendulum coupling $g\gg \kappa$. The choice of parameters is similar to the recently realized circuit with the same geometry~\cite{pietikainen2017}.

The transition energies of the transmon are determined by the Josephson energy $E_{\rm J}$ and by the charging energy $E_{\rm C}$, which can be adjusted by the design of the shunting capacitor $C_{\rm B}$, see Fig.~\ref{fig:oscpendevals}. The transition energy between the lowest two energy eigenstates is given by $\hbar\omega_{\rm q} \approx \sqrt{8E_{\rm J}E_{\rm C}}-E_{\rm C} = E_{\rm C} (\sqrt{8\eta}-1)$. We will study the onset of the nonlinearities for different 
drive detunings 
$\delta_{\rm d}=\omega_{\rm d}-\omega_{\rm r}$ as a function of the drive amplitude $A$. We are especially interested in the resonant case $\delta_{\rm q}=\omega_{\rm q}-\omega_{\rm r}=0$. 
The detuned case has been previously studied in more detail in 
Refs.~\cite{pietikainen2017,pietikainen2018,lescanne2018,verney2018}. 
We have used a temperature value of $k_{\rm B}T/(\hbar \omega_{\rm r})=0.13$ which corresponds to $T\approx 30$ mK for a transmon with $\omega_{\rm q}/(2\pi) = 5$ GHz. 

\section{Numerical results}\label{sec:IV}

\subsection{Classical system}

Classically, we can understand the behaviour of our system as follows: the pendulum-resonator forms a coupled system, whose normal modes can be obtained. However, because the pendulum is nonlinear, the normal-mode frequencies of the coupled-system depend on the oscillation amplitude of the pendulum. The resonator acts also as a filter for the drive, which is thus applied to the pendulum. As the oscillation amplitude of the pendulum increases, the normal-mode frequency shifts, an effect which is responsible for photon blockade. Eventually the pendulum reaches the free rotor regime, where the Josephson energy becomes negligible. As a consequence, the nonlinearity no longer plays any role, and the resulting eigenmode of the system is that of the bare resonator.

\begin{figure}[ht!]
\includegraphics[width=1.0\linewidth]{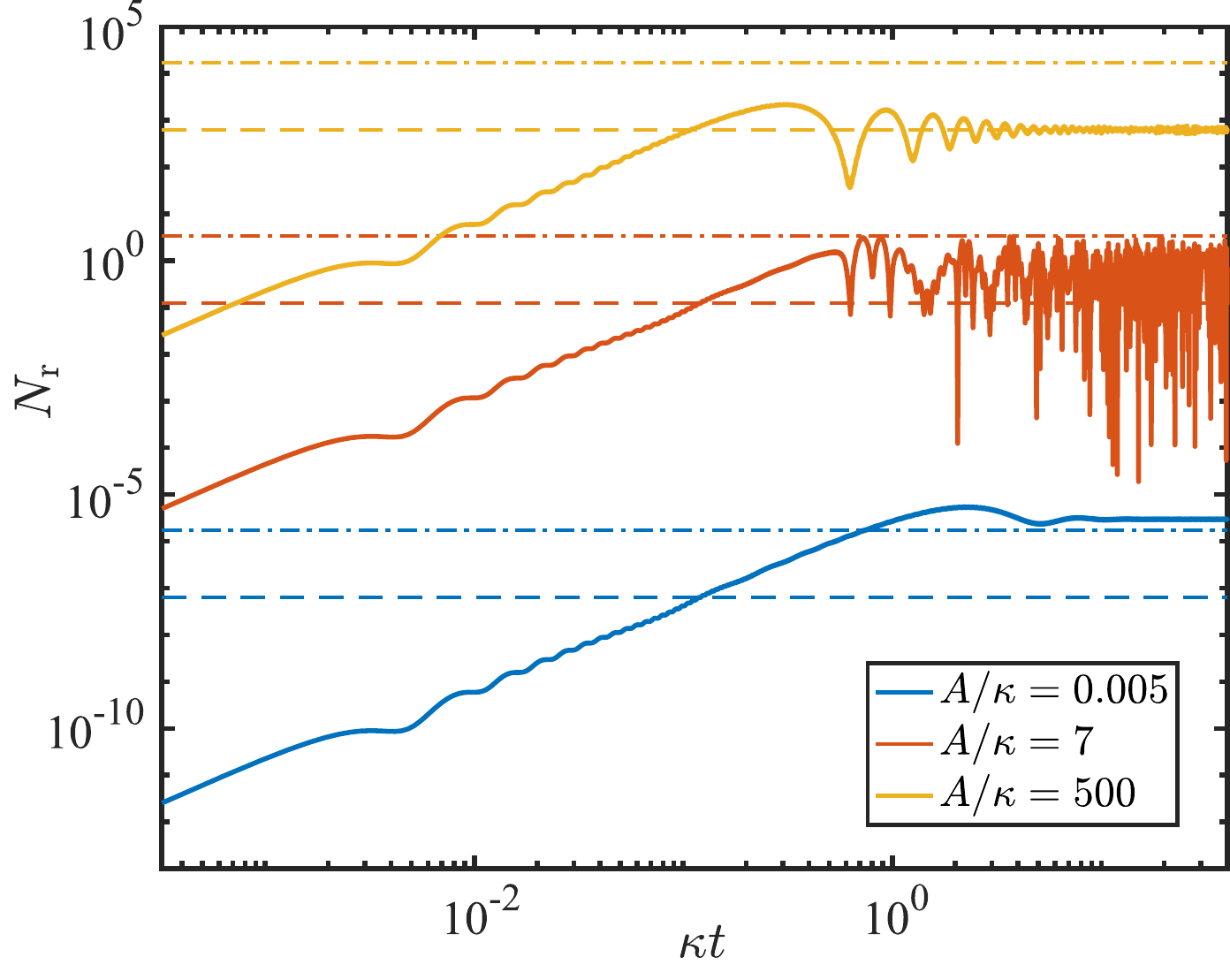}\hfill
\caption{Classical dynamics of the resonator occupation $N_{\rm r}$ of the driven and dissipative resonator-transmon system. We show data for linear ($A/\kappa = 0.005$, \added{bottom line}), chaotic ($A/\kappa = 7$, \added{middle line}), and the bare-resonator ($A/\kappa = 500$, \added{top line}) regimes. The bare-oscillator occupation in the steady state is given by Eq.~(\ref{eq:anocc}) and indicated with dashed lines. We also show with dot-dashed lines the steady-state photon numbers for the linearized system, as given in Eq.~(\ref{eq:linearNr}). 
We have used the pendulum dissipation rate $\gamma/\omega_{\rm r} =2\times 10^{-4}$. The other parameters are listed in Table~\ref{tab:params1}.} \label{fig:classsteps}
\end{figure}

We first solve the classical equation of motion (see Appendix~\ref{app:classeom}) for the driven and damped resonator-transmon system. We study the steady-state occupation $N_{\rm r}$ of the resonator as a function of the drive amplitude. Classically, one expects that the coupling to the transmon causes deviations from the bare resonator occupation
\begin{equation}\label{eq:anocc}
N_{\rm bare} = \frac14\frac{A^2}{\delta_{\rm d}^2 + \kappa^2/4}.
\end{equation}
We emphasize that $N_{\rm bare} \leq A^2/\kappa^2$ where the equality is obtained if the drive is in resonance, i.e. if $\delta_{\rm d}=0$. The numerical data for $\delta_{\rm d}/\omega_{\rm r} =-0.02$ is shown in Fig.~\ref{fig:classsteps}. We compare the numerical data against the bare-resonator photon number in Eq.~(\ref{eq:anocc}), and against the photon number of the linearized system, see Appendix~\ref{app:classeom},
\begin{equation}\label{eq:linearNr}
N_{\rm lin} = \frac{A^2}{4}\frac{1}{\left(\delta_{\rm d}-g_{\rm eff}^2\frac{\delta_{\rm p}}{\delta_{\rm p}^2+\gamma^2/4}\right)^2+\left(\frac{\kappa}{2}+g_{\rm eff}^2\frac{\gamma/2}{\delta_{\rm p}^2+\gamma^2/4}\right)^2},
\end{equation}
where $\delta_{\rm p} = \omega_{\rm d}-\omega_{\rm p}$, $\hbar \omega_{\rm p} = \sqrt{8E_{\rm J}E_{\rm C}}$, $g_{\rm eff} = g\sqrt[4]{\eta/32}$, and $\gamma$ is the dissipation rate of the pendulum. The above result is obtained by linearizing the pendulum potential which results to system that is equivalent to two coupled harmonic oscillators. We find in Fig.~\ref{fig:classsteps} that for small drive amplitude $A/\kappa = 0.005$, the steady state of the resonator photon number is given by that of the linearized system. As a consequence, both degrees of freedom oscillate at the drive frequency and the system is classically stable. The small deviation between the numerical and analytic steady-state values is caused by the rotating-wave approximations that were made for the coupling and the drive in the derivation of Eq.~(\ref{eq:linearNr}). 

If the drive amplitude is increased to $A/\kappa =7$, the nonlinearities caused by the cosinusoidal Josephson potential generate chaotic behavior in the pendulum. As a consequence, the photon number does not find a steady state but, instead, displays aperiodic chaotic oscillations around some fixed value between those of the bare resonator and the linearized system. This value can be found by studying the long-time average in the steady state. For very high drive amplitude $A/\kappa = 500$, the photon number in the classical system is given by that of the bare resonator in Eq.~(\ref{eq:anocc}). Physically this means that for strong driving, the pendulum experiences rapid free rotations and, as a consequence, its contribution to the photon dynamics is zero on average. 

In Fig.~\ref{fig:Occupation7}(a), we study in more detail how the classical steady-state photon number of the resonator changes as the coupled system goes through the transition between the linearized oscillations in the weak driving regime and the bare-resonator oscillations for strong driving. In the absence of driving the steady-state photon number is zero in accordance to Eq.~(\ref{eq:linearNr}). For low drive amplitudes, the resonator-transmon system can be approximated as a driven and damped Duffing oscillator. We show in Appendix~\ref{app:classeom} that the system has one stable solution for drive amplitudes $A<A_{\rm min}$ and $A>A_{\rm crit}$, and two stable solutions for $A_{\rm min}<A<A_{\rm crit}$ where
\begin{eqnarray}
A_{\rm min} &=& \tilde\gamma\sqrt{2(\tilde\omega_{\rm p}^2-\omega_{\rm d}^2)}\frac{\sqrt{(\tilde{\omega}_{\rm r}^2-\omega_{\rm d}^2)^2+\kappa^2\omega_{\rm d}^2}}{g\omega_{\rm r}\omega_{\rm p}},\label{eq:duffminimal}\\
A_{\rm crit} &=& \sqrt{\frac{8}{27}}\sqrt{(\tilde{\omega}_{\rm p}^2-\omega_{\rm d}^2)^3}\frac{\sqrt{(\tilde{\omega}_{\rm r}^2-\omega_{\rm d}^2)^2+\kappa^2\omega_{\rm d}^2}}{g\omega_{\rm r}\omega_{\rm d}\omega_{\rm p}},\label{eq:duffan}
\end{eqnarray}
where we have defined the renormalized oscillator frequency and transmon dissipation rate as $\tilde{\omega}_{\rm r}^2 = \omega_{\rm r}^2 - g^2\hbar \omega_{\rm r}/(4E_{\rm C})$ and $\tilde{\gamma} = \gamma+gg_1\kappa\omega_{\rm d}^2/[(\tilde{\omega}_{\rm r}^2-\omega_{\rm d}^2)^2+\kappa^2\omega_{\rm d}^2]$, respectively, the classical oscillation frequency of the linearized transmon as $\hbar\omega_{\rm p}=\sqrt{8E_{\rm J}E_{\rm C}}$, and the renormalized linearized transmon frequency as $\tilde{\omega}_{\rm p}^2 = \omega_{\rm p}^2-g^2 \omega_{\rm d}^2/(\tilde{\omega}_{\rm r}^2-\omega_{\rm d}^2)[\hbar \omega_{\rm r}/(4E_{\rm C})]$.

For amplitudes $A<A_{\rm min}$, the classical system behaves as a two-oscillator system 
and the photon number has the typical quadratic dependence on the drive amplitude in Eq.~(\ref{eq:linearNr}). As the drive amplitude becomes larger than $A_{\rm min}$ deviations from the linearized model emerge. In addition, the system becomes bistable. If $A \approx A_{\rm crit}$, the number of stable solutions for the Duffing oscillator is reduced from two to one. This is displayed by the abrupt step in the photon number of the classical solution in Fig.~\ref{fig:Occupation7} around $A_{\rm crit}/\kappa =1.2$. The remaining high-amplitude stable solution appears as a plateau which reaches up to the drive amplitude $A/\kappa \approx 5.6$. If the drive amplitude is further increased, the higher order terms beyond the Duffing approximation render the motion of the classical system chaotic, as described already in Fig.~\ref{fig:classsteps}. For large drives, the classical photon number approaches asymptotically the photon number of the bare resonator.

\subsection{Quantum description}\label{sec:quantdesc}

The transition between the motion of linearized and bare-resonator oscillations is characteristic to oscillator-pendulum systems. However, we show here that in the quantum mechanical context, the onset of the nonlinear dynamical behaviour turns out to be quantitatively different from that provided by the above classical model. 
This was also observed in recent experimental realization with superconducting circuits~\cite{pietikainen2017}. 

\begin{figure}[t!]
  \centering
\includegraphics[width=1.0\linewidth]{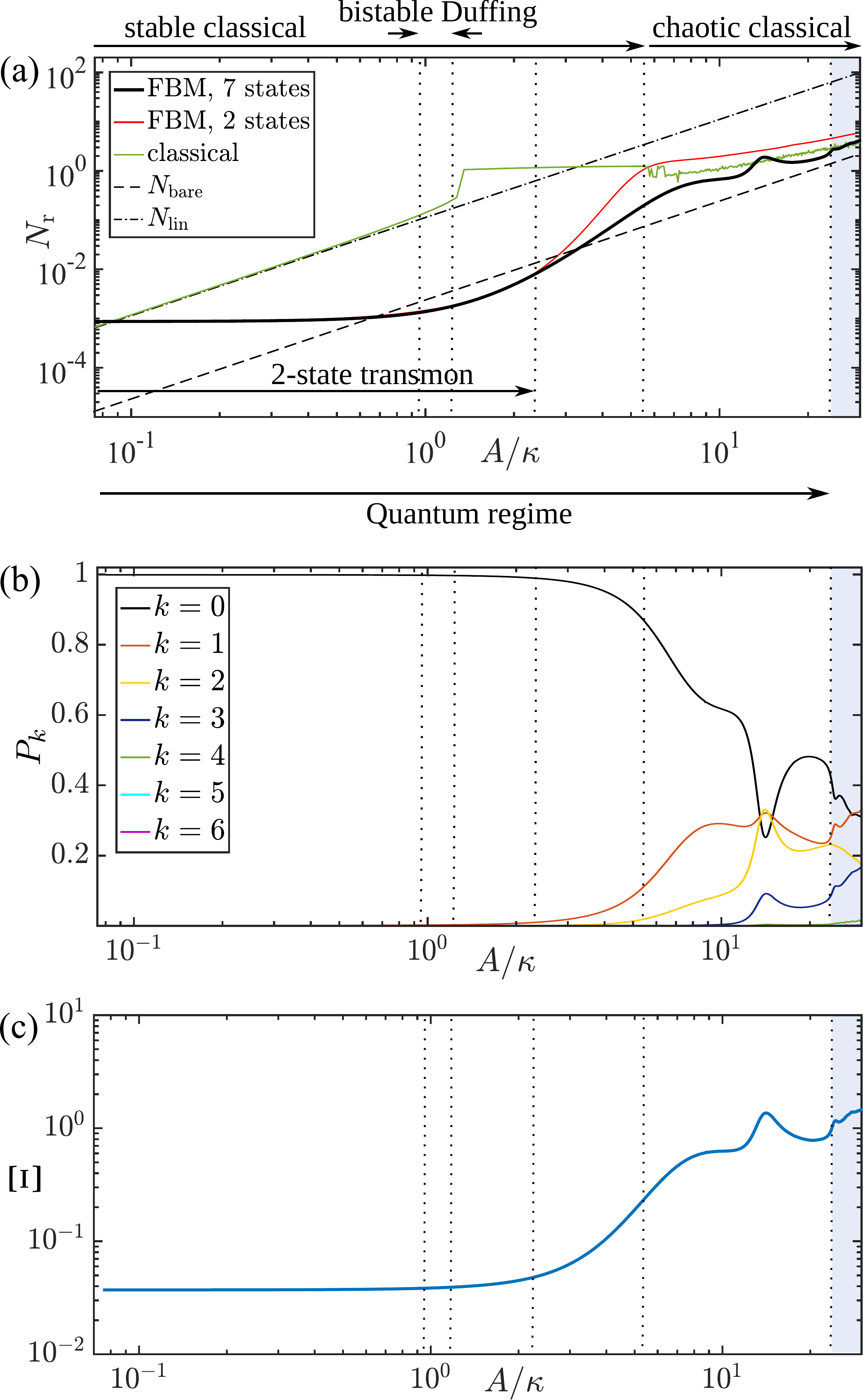}
\caption{Onset of the nonlinearities in the driven system. (a) The steady-state photon number $N_{\rm r}$ as a function of the drive amplitude. We compare the Floquet--Born--Markov (FBM) simulation with seven transmon states against the corresponding solutions for the Rabi Hamiltonian and the classical system. The classical region of bistability occurs between $A_{\rm min}/\kappa = 0.97$ and $A_{\rm crit}/\kappa=1.2$, given by Eqs.~(\ref{eq:duffminimal}) and~(\ref{eq:duffan}), respectively. The classical simulation demonstrates switching between the two stable solutions at $A\approx A_{\rm crit}$. 
We also show the photon numbers of the linearized system and the bare resonator, as given by Eqs.~(\ref{eq:anocc}) and (\ref{eq:linearNr}), respectively. Note that both axes are shown in logarithmic scale.
(b) Occupation probabilities $P_k$ of the transmon eigenstates calculated using FBM. We indicate the regime of classical response with the shaded region in both figures. \added{Occupation probabilities for states with $k\geq 4$ are negligible for the used parameters.}
(c) Order parameter $\Xi$ defined in Eq.~(\ref{eq:orderparameter}).
We have used $n_{\rm g}=0$ and the drive detuning $\delta_{\rm d}/\omega_{\rm r} = -0.02$. Other parameters are listed in  Table~\ref{tab:params1}.}\label{fig:Occupation7}
\end{figure}

In the quantum-mechanical treatment, we calculate the steady-state photon number in the resonator as a function of the drive amplitude using the Floquet--Born--Markov master equation presented in Sec.~\ref{sec:FBM}. 
We have confirmed that for the used values of the drive amplitude the simulation has converged for the truncation of seven transmon states and 60 resonator states. We compare the quantum results against those given by the classical equation of motion and study also deviations from the results obtained with the two-state truncation of the transmon.  In Fig.~\ref{fig:Occupation7}, we present the results corresponding to gate charge $n_{\rm g}=0$,  where the resonator, the transmon, and the drive are nearly resonant at low drive amplitudes. The used parameters are the same as in Fig.~\ref{fig:classsteps} and listed in Table~\ref{tab:params1}. 

First, we notice in Fig.~\ref{fig:Occupation7}(a) that even in the absence of driving there always exists a finite photon occupation of $N_{\rm r} \approx 10^{-3}$  in the ground state, contrary to the classical solution which approaches zero. At zero temperature, the existence of these ground-state photons~\cite{lolli2015} originates from the terms in the interaction Hamiltonian that do not conserve the number of excitations and are neglected in the rotating-wave approximation resulting in Eq.~(\ref{eq:ManyStatesHam_simple}).  For the two-state truncation of the transmon, one can derive a simple analytic result for the ground-state photon number by treating these terms as a small perturbation. In the second order in the perturbation parameter $g/\omega_{\rm r}$, one obtains that the number of ground-state photons is given by $N_{\rm r} \approx (g/2\omega_{\rm r})^2$.  We have confirmed that our simulated photon number at zero driving is in accordance with this analytic result if $T=0$ and $g/\omega_{\rm r}\ll 1$. The photon number at zero driving obtained in Fig.~\ref{fig:Occupation7}(a) is slightly higher due to additional thermal excitation - in the simulations we use a finite value for temperature, see Table~\ref{tab:params1}. 

As was discussed in the previous section, the resonator photon number of a classical system increases quadratically with the drive amplitude. For amplitudes $A<A_{\rm crit}$, the classical system can be approximated with a linearized model formed by two coupled harmonic oscillators.  However, in the quantum case the energy levels are discrete and, thus, the system responds only to a drive which is close to resonance with one of the transitions. In addition, the energy levels have non-equidistant separations which leads to a reduction of the photon number compared to the corresponding classical case, referred to as the photon blockade. This is also apparent in Fig.~\ref{fig:Occupation7}(a).

We emphasize that 
the photon-blockade is  quantitatively strongly-dependent on the transmon truncation. This can be seen as the deviation between the two and seven state truncation results for $A/\kappa>1$ in Fig.~\ref{fig:Occupation7}(a). We further demonstrate this by showing the transmon occupations $P_{\rm k}$ in Fig.~\ref{fig:Occupation7}(b). For weak drive amplitudes, the transmon stays in its ground state. The 
excitation of the two-level system is accompanied by excitations of the transmon to several its bound states
. If $A/\kappa \geq 30$, the transmon escapes its potential well and also the free rotor states start to gain a finite occupation. This can be interpreted as a transition between the Duffing oscillator and free rotor limits of the transmon, see Appendix \ref{app:eigenvalue}.  As a consequence, the response of the quantum system resembles its classical counterpart. We will study the photon blockade in more detail in the following section.

\subsubsection*{Order parameter}

In order to characterize the transition between the quantum and classical regime, we can also study the behaviour of the order parameter $\Xi$ defined as the expectation value of the coupling part of the Hamiltonian in Eq.~(\ref{eq:ManyStatesHam}), normalized with $\hbar g$, as
\begin{equation}\label{eq:orderparameter}
\Xi = \left|\left\langle (\hat a^{\dag} + \hat a)\sum_{k,\ell}\hat \Pi_{k,\ell}\right\rangle\right|,
\end{equation}
previously introduced and used for the off-resonant case in Ref.~\cite{pietikainen2017}. To get an understanding of its behavior, let us evaluate it for the resonant Jaynes--Cummings model, 
\begin{equation}\label{eq:orderparameterJC}
\Xi_{\rm JC} = \left|\left\langle n_{r}, \pm| (\hat a^{\dag} + \hat a)\sigma_{x} |n_{r},\pm \right\rangle \right| = \sqrt{n_{r}},
\end{equation}
therefore it correctly estimates 
the absolute value of the cavity field operator. At the same time, when applied to the full Rabi model, it includes the effect of the 
terms that do not conserve the excitation number.

In Fig.~\ref{fig:Occupation7}(c), we present $\Xi$ as a function of 
the drive amplitude $A$. Much like in the off-resonant case, this order parameter displays a marqued increase by one order of magnitude across the transition region. 

\subsection{Photon blockade: dependence on the drive frequency}

\begin{figure*}[ht!]
\includegraphics[width=1.0\linewidth]{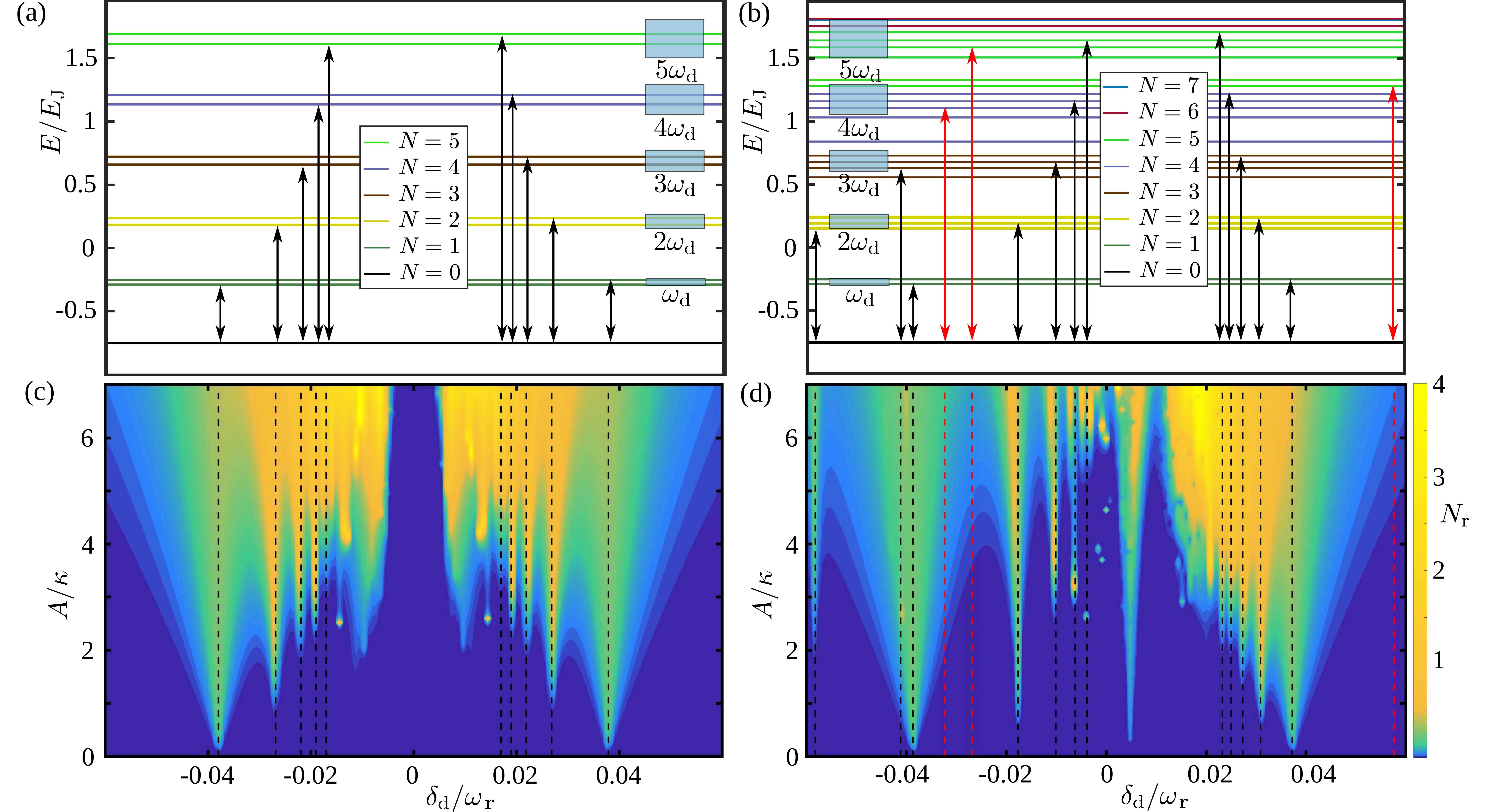}
\caption{
Steady-state photon number as a function of drive detuning and amplitude for the resonant case with $\delta_{\rm q}= 0.0$. Other parameters used in the simulation are listed in the Table~\ref{tab:params1}. We present the energy level diagrams of the coupled pendulum-resonator for the (a) two-state and (b) seven-state truncation for the transmon, and the corresponding simulations (c) and (d) for the average number of photons $N_{\rm r}$. 
In the diagrams (a) and (b), the states are labeled by the excitation number $N$ which is a good quantum number if the rotating-wave approximation is valid. We also highlight by blue \added{(gray)} rectangles the range of energies accessible by 1,2,3,4, and 5 - photon transitions, corresponding to the range of drive frequencies $\delta_{\rm d}/\omega_{\rm r} \in\{-0.06, 0.06 \}$ in (c) and (d). The vertically aligned arrows indicate the locations of transitions that correspond to the multi-photon blockades in (c) and (d), which are denoted with dashed lines. Absent transitions are denoted in red \added{(gray)}.}\label{fig:DriveDetuning}
\end{figure*}

Here, we discuss in more detail the phenomenon of photon blockade in the pendulum-resonator system as a function of the drive detuning $\delta_{\rm d} = \omega_{\rm d}-\omega_{\rm r}$.
First, we consider the transition between the ground state and the state $|n_{r},\pm\rangle$ [Eq.~(\ref{eq:JCstates})] of the resonant Jaynes--Cummings system ($\omega_{\rm q}=\omega_{\rm r}$). We recall that the selection rules Eqs.~(\ref{eq:selection1}) and (\ref{eq:selection2}) allow only direct transitions that change the excitation number by one. 
However, at higher amplitudes the probability of higher order processes is no longer negligible and excited states can be populated by virtual non-resonant single-photon transitions. 
As a consequence, one obtains the resonance condition for multi-photon transitions as $n_{r}\omega_{\rm d} =n_{r}\omega_{\rm r} \pm \sqrt{n_{r}}g$. Because the energy-level structure is non-equidistant, the drive couples only weakly to other transitions in the system. In the absence of dissipation, the dynamics of the Jaynes--Cummings system can, thus, be approximated in a subspace spanned by the states $\{|0,0\rangle,|n_{r},\pm\rangle\}$.

Thus, one expects that, due to the driving, the system goes through $n_{r}$-photon Rabi oscillations between the basis states of the subspace. The Rabi frequency $\Omega_{n_{r},\pm}$ of such process is proportional to the corresponding matrix element of the driving Hamiltonian in Eq.~(\ref{eq:Hd}) and the drive amplitude $A$. Consequently, the time-averaged photon number in the system is $N_{\rm r} = (n_{r}-\frac{1}{2})/2$. 
The driving does not, however, lead into a further increase of the photon number either because the drive is not resonant with transitions from the state $|n_{r},\pm\rangle$ to higher excited states or the matrix element of the resonant transitions are negligibly small. We are referring to this phenomenon as $n_{r}$-photon blockade. 

Dissipation modifies somewhat this picture, as it causes transitions outside the resonantly driven subspace. As a consequence, the average photon number decays with a rate which is proportional to $\kappa$. Thus, the steady state of such system is determined by the competition between the excitation and relaxation processes caused by the drive and the dissipation, respectively. At low temperatures, the occupation in the ground state becomes more pronounced as the dissipation causes mostly downward transitions. Thus, the steady-state photon number is reduced compared to the time-averaged result for Rabi-driven non-dissipative transition. 
This was visible already in Fig.~\ref{fig:Occupation7} in which the data was obtained with the two-state truncation and corresponds to the 4-photon blockade of the Jaynes--Cummings system. 

The diagram in Fig.~\ref{fig:DriveDetuning}(a) represents the eigenenergies of the Hamiltonian for Eq.~(\ref{eq:ManyStatesHam}) in the two-state truncation for the transmon. The states are classified according to the excitation number $N$ from Eq.~(\ref{eq:exitationN_2}). We note that, here, we do not make a rotating-wave approximation and strictly speaking $N$ is, therefore, not a good quantum number. However, it still provides a useful classification of the states since the coupling frequency is relatively small, i.e. $g/\omega_{\rm r}=0.04$. 

In Fig.~\ref{fig:DriveDetuning}(c), we show the photon blockade spectrum of the resonator-transmon system as a function of the drive detuning $\delta_{\rm d}$, obtained numerically with the Floquet--Born--Markov master equation. 
Here, one can clearly identify the one-photon blockade at the locations where the drive frequency is in resonance with the single-photon transition frequency of the resonator-transmon system~\cite{bishop2009}, i.e. when $\delta_{\rm d}= \pm g$. Two, three, and higher-order blockades occur at smaller detunings and higher drive amplitudes, similar to Ref.~\cite{carmichael2015}. Transitions involving up to five drive photons are denoted in the diagram in Fig.~\ref{fig:DriveDetuning}(a) and are vertically aligned with the corresponding blockades in Fig.~\ref{fig:DriveDetuning}(c).
At zero detuning, there is no excitation as the coupling to the transmon shifts the energy levels of the resonator so that there is no transition corresponding with the energy $\hbar\omega_{\rm r}$.
We also note that the photon-number spectrum is symmetric with respect to the drive detuning $\delta_{\rm d}$. We see this same symmetry also in Eq.~(\ref{eq:linearNr}) for the linearized classical system when the classical linearized frequency of the transmon is in resonance with the resonator frequency, i.e. when $\omega_{\rm p}=\omega_{\rm r}$.

However, in experimentally relevant realisations of such systems the higher excited states have a considerable quantitative influence to the photon-number spectrum. We demonstrate this by showing data for the seven-state transmon truncation in Figs.~\ref{fig:DriveDetuning}(b) and (d). 
The eigenenergies shown in Fig.~\ref{fig:DriveDetuning}(b) are those obtained in Fig.~\ref{fig:oscpendevals} at resonance ($\omega_{\rm r}=\omega_{\rm q}$). We have again confirmed that for our choice of drive amplitudes and other parameters, this truncation is sufficient to obtain converged results with the Floquet--Born--Markov master equation. We observe that the inclusion of the higher excited states changes considerably the observed photon number spectrum. However, the states can again be labeled by the excitation number $N$ which we have confirmed by numerically calculating $N=\langle \hat N\rangle$ for all states shown in Fig.~\ref{fig:DriveDetuning}(c). The relative difference from whole integers is less than one percent for each shown state. 
Corresponding to each $N$, the energy diagram forms blocks containing $N+1$ eigenstates with (nearly) the same excitation number, similar to the doublet structure of the Jaynes--Cummings model. Contrary to the two-state case, these blocks start to overlap if $N>4$ for our set of parameters, as can be seen in Fig.~\ref{fig:DriveDetuning}(b).

The number of transitions that are visible for our range of drive frequencies and amplitudes in Fig.~\ref{fig:DriveDetuning}(d) is, thus, increased from ten observed in the Jaynes--Cummings case to 15 in the seven-state system. However, some of these transitions are not visible for our range amplitudes due to the fact that the corresponding virtual one-photon transitions are not resonant and/or have small transition matrix elements. In addition, the spectrum is asymmetric with respect to the detuning as the multi-photon resonances are shifted towards larger values of $\delta_{\rm d}$. As a consequence, the break-down of the photon blockade at $\delta_{\rm d}=0$ occurs at much lower amplitudes as is observed in the Jaynes--Cummings system~\cite{carmichael2015}.

\subsection{Approaching the ultrastrong coupling}

\begin{figure}[h!]
\includegraphics[width=1.0\linewidth]{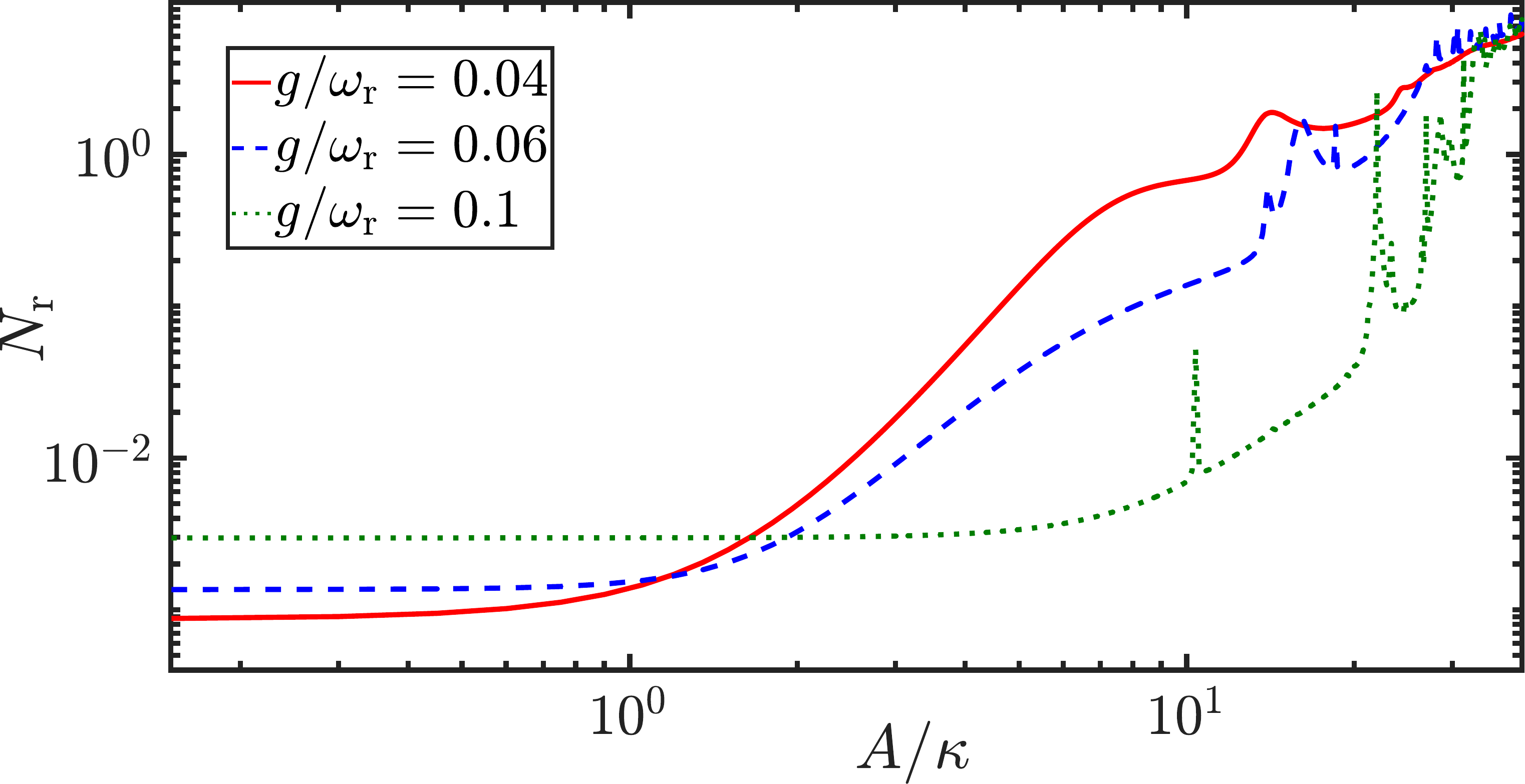}\hfill
\caption{
Steady-state photon number $N_{\rm r}$ \added{of the multilevel Josephson--Rabi model} as a function of drive amplitude for different coupling strengths. The simulations are realized using the Floquet--Born--Markov approach with the seven-state truncation for the transmon. 
The drive detuning is $\delta_{\rm d}/\omega_{\rm r} = -0.02$ and also the other 
parameters are the same as in Table~\ref{tab:params1}.}\label{fig:ultrastrong}
\end{figure}

For most applications in quantum information processing a relative coupling strength $g/\omega_{\rm r}$ of a few percent 
is sufficient. However, recent experiments with superconducting circuits have demonstrated that it is possible to increase this coupling into the ultrastrong regime ($g/\omega_{\rm r} \sim 0.1 - 1$) and even further in the deep strong coupling regime ($g/\omega_{\rm r} \geq 1$) \cite{FornDiaz2018, Gu2017, Kockum2019}. 
While the highest values have been obtained so far with flux qubits, vacuum-gap transmon devices with a similar electrical circuit as in Fig.~\ref{fig:oscpendevals}(a) can reach
$g/\omega_{\rm r} = 0.07$~\cite{Bosman2017a} and $g/\omega_{\rm r} = 0.19$~\cite{Bosman2017b}. \added{To study the multilevel Josephson--Rabi model at higher couplings, we present in Fig.~\ref{fig:ultrastrong} results} for the average number of photons in the resonator for couplings  $g/\omega_{\rm r} = 0.04, 0.06$, and $0.1$, employing the Floquet--Born--Markov approach to dissipation.

At low drive powers the two-level approximation can be used for the transmon, and the Josephson pendulum-resonator system maps into the quantum Rabi model. From Fig.~\ref{fig:ultrastrong} we see that the average number of photons $N_{\rm r}$ in the resonator is not zero even in the ground state; this number clearly increases as the coupling gets stronger. As noted also before, this is indeed a feature of the quantum Rabi physics: differently from the Jaynes--Cummings model where the ground state contains zero photons, the terms that do not conserve the excitation number 
in $\hat{H}_{\rm c}$ lead to a ground state which is a superposition of transmon 
and resonator 
states with non-zero number of excitations. \added{Similar to Sec.~\ref{sec:quantdesc},} the perturbative formula $N_{\rm r} \approx (g/2\omega_{\rm r})^2$ \deleted{again} 
approximates very well the average number of photons at zero temperature, while in Fig.~\ref{fig:ultrastrong} we observe slightly higher values due to the  finite temperature. As the drive increases, we observe that the photon blockade tends to be more effective for large $g$'s. Interestingly, the transition to a classical state also occurs more abruptly as the coupling gets stronger. We have checked that this coincides with many of the upper levels of the transmon being rapidly populated. Due to this effect, the truncation to seven states (which is the maximum that our code can handle in a reasonable amount of time) becomes less reliable and artefacts such as the sharp resonances at some values start to appear. 

\subsection{\added{Correlation functions}}
\begin{figure*}[ht!]
\includegraphics[width=0.9\linewidth]{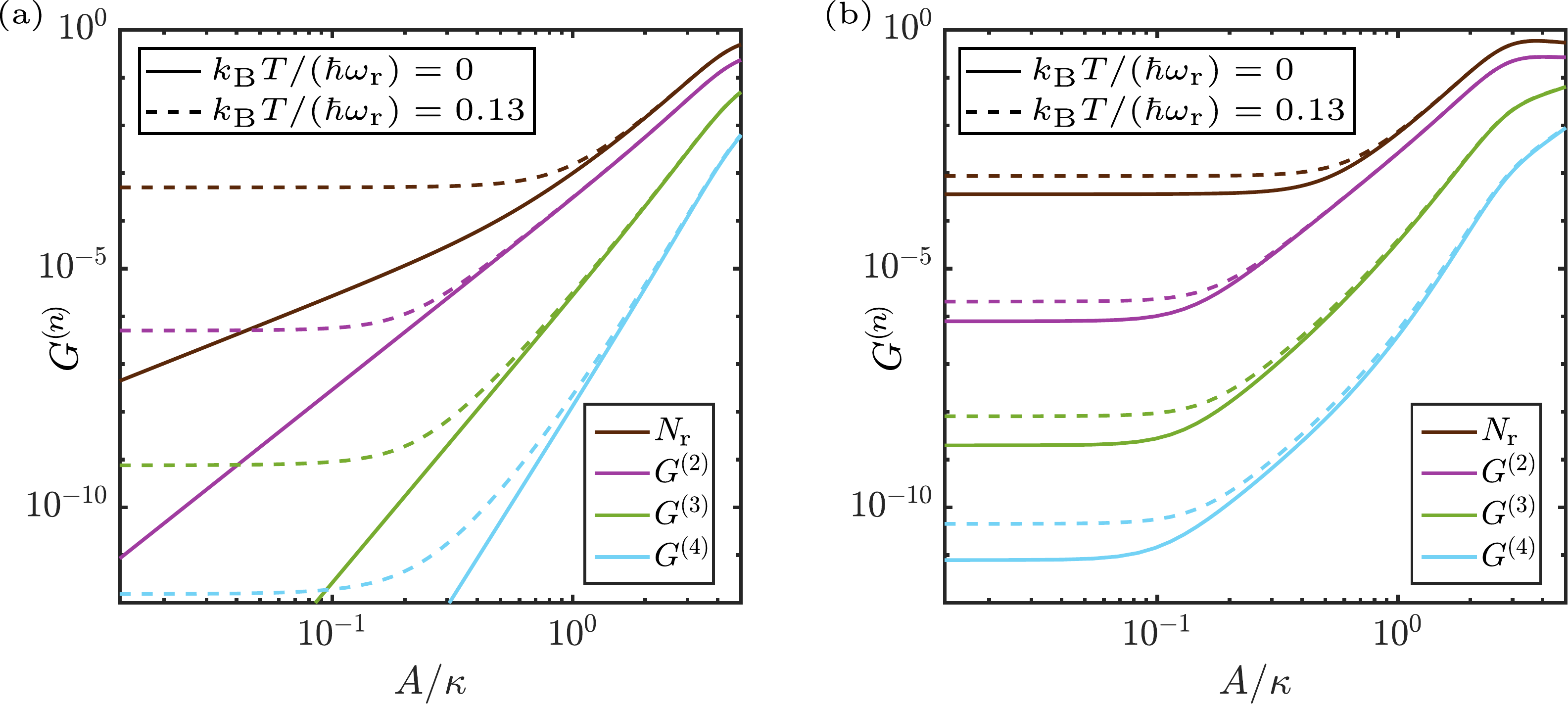}\hfill
\caption{\added{Correlations $G^{(n)}$ as a function of the drive amplitude. The system is driven at the 2-photon blockade frequency $\delta_{\rm d}/\omega_{\rm r} = -0.018$. The calculations are done for a seven-state transmon (a) by performing the rotating-wave approximation, and (b) with the full Josephson--Rabi Hamiltonian. The solid lines correspond to zero temperature and the dashed lines to $k_{\rm B}T/(\hbar\omega_{\rm r})=0.13$, while the other
parameters are the same as in Table~\ref{tab:params1}.}}\label{fig:Gn}
\end{figure*}

\added{Photon correlations provide the essential information about the statistics of the quantum field leaking out of the cavity~\cite{Glauber1963, Bozyigit2010, Hamsen2017}. Especially interesting is the zero delay correlation function~\cite{Fink2018, Bin2018}.
The $n$th order correlation function is defined as
\begin{equation}\label{eq:Gn}
G^{(n)} = \langle \hat{a}^{\dagger n} \hat{a}^n \rangle \,,
\end{equation}
where $G^{(1)}=N_{\rm r}$. 
In Fig.~\ref{fig:Gn} we have calculated numerically the correlation functions for the seven state transmon (a) with the rotating-wave approximation and (b) including the counter-rotating (Bloch--Siegert) terms present in the Josephson--Rabi model. }

\added{
At zero driving and zero temperature, the system is in its ground state. If the rotating-wave approximation is used, this state is $\vert 0,0 \rangle$. This means that all $G^{(n)} = 0$. This can be seen in Fig.~\ref{fig:Gn}(a). Without the rotating-wave approximation, the analytic form of the ground state has not been found even for the two-level Rabi Hamiltonian. We obtain an approximative ground state of the Rabi Hamiltonian by treating the counter-rotating terms $\propto (\hat a^{\dag}\hat \sigma_+ + \hat a \hat \sigma_-)$ as a small perturbation. We obtain in the second order of the perturbation parameter $g_{01}/\omega_{\rm r}$ that the non-normalized ground state can be written as 
\begin{equation}
\vert g\rangle \approx \vert 0,0\rangle -\frac{g_{01}\omega_{\rm r}}{2\omega_{\rm r}^2 -g_{01}^2} \vert 1,1\rangle
+\frac{\sqrt{2}g_{01}^2}{2(2\omega_{\rm r}^2 -g_{01}^2)} \vert 2,0\rangle \,.
\end{equation}
This gives $G^{(1)} = 3.7 \times 10^{-4}$ and $G^{(2)} = 5.2 \times 10^{-7}$ with the parameters used here. The corresponding values with seven state Josephson--Rabi Hamiltonian from Fig.~\ref{fig:Gn}(b) are $G^{(1)} = 3.6 \times 10^{-4}$ and $G^{(2)} = 7.9 \times 10^{-7}$, demonstrating very good agreement.
As seen from Fig.~\ref{fig:Gn}, raising the temperature reduces the difference between the two cases, since now the system contains thermally-activated photons even at zero drive amplitude.}
\added{Overall, as expected, in the quantum regime the $G^{(n)}$'s are very small, but once the system approaches the classical regime as $A/\kappa \gg 1$ the values of $G^{(n)}$ become larger than one. The marked difference between the rotating-wave approximation result shown in Fig.~\ref{fig:Gn}(a) and that of the full Josephson-Rabi model in Fig.~\ref{fig:Gn}(b) suggest the use of statistics as a detection method for the ultrastrong-coupling regime. 
}

\subsection{Dependence on the gate charge}

\begin{figure}
  \centering
\includegraphics[width=\linewidth]{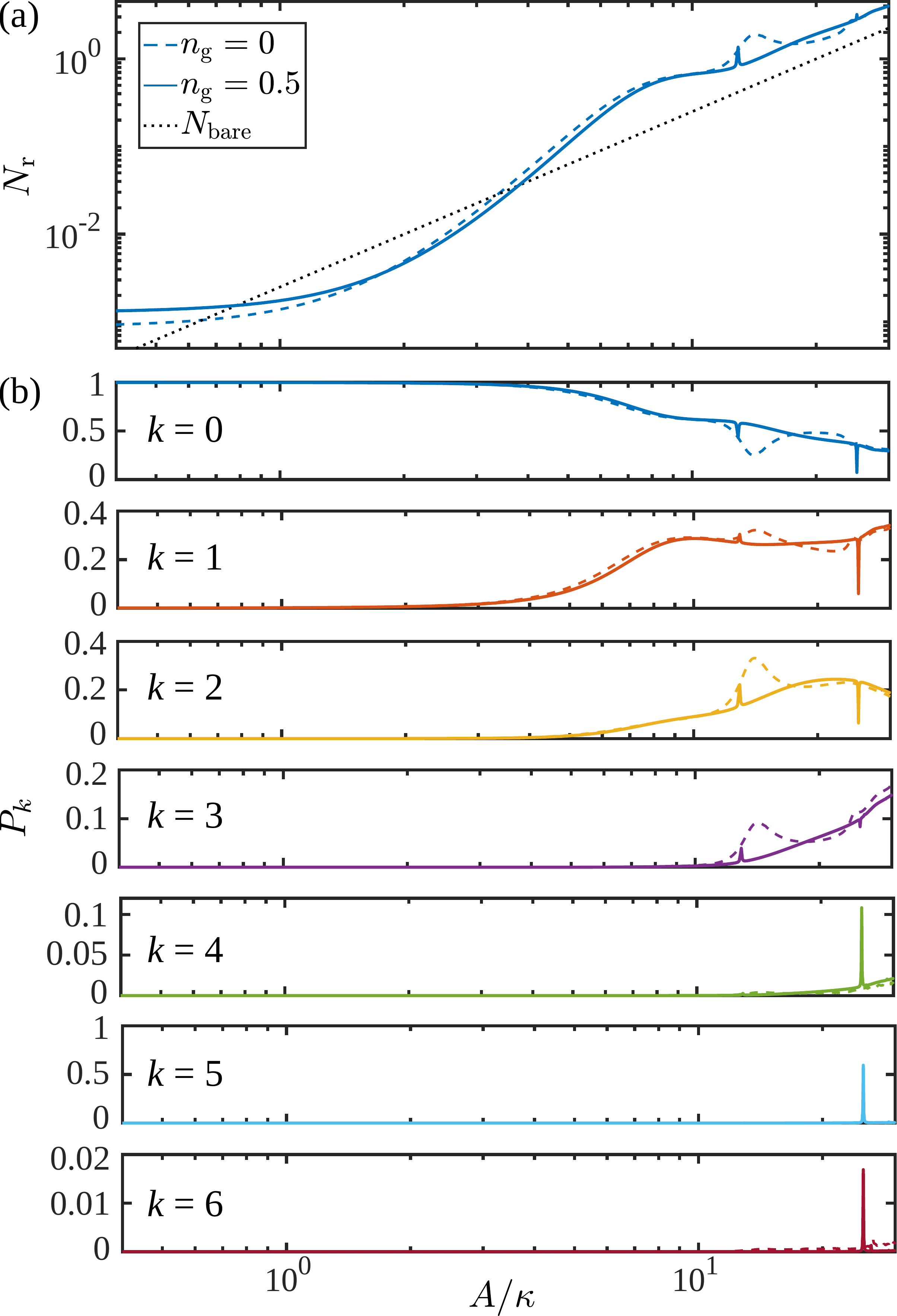}
\caption{
Onset of nonlinearity as a function of the gate charge. (a) The photon number $N_{\rm r}$ as a function of the drive amplitude. We compare the numerical data for $n_{\rm g} = 0$ and $n_{\rm g} = 0.5$ obtained with seven transmon states. 
(b) Corresponding occupations $P_{k}$ in the transmon eigenstates. 
The drive detuning is $\delta_{\rm d}/\omega_{\rm r} = -0.02$ and also the other 
parameters are the same as in Table~\ref{tab:params1}.}\label{fig:Ng}
\end{figure}

If the transmon is only weakly nonlinear, i.e. $\eta \gg 1$, its lowest bound eigenstates are insensitive to the gate charge, see Appendix \ref{app:eigenvalue}. 
As a consequence, one expects that the value of the gate charge should not affect the photon-number response to a weak drive. However, as the amplitude of the drive is increased, the higher excited states of the transmon become occupied, as discussed in the context of Fig.~\ref{fig:Occupation7}. Especially, the transition region between the quantum and classical responses 
should be dependent on the gate-charge dispersion of the transmon states. 
We demonstrate this in Fig.~\ref{fig:Ng}, where we show the simulation data for the gate-charge values $n_{\rm g}=0$ and $n_{\rm g}=0.5$. Clearly, in the weak driving regime, the responses for the two gate-charge values are nearly equal. The deviation in the photon number is of the order of $10^{-3}$, which is explained by our rather modest value of $\eta=30$. 

The deviations between the photon numbers of the two gate-charge values are notable if $A/\kappa = 10\ldots 20$. In this regime, the transmon escapes the subspace spanned by the two lowest eigenstates and, thus, the solutions obtained with different gate-charge numbers are expected to differ. At very high amplitudes, the free-rotor states with $k\geq 6$ also begin to contribute the dynamics. These states have a considerable gate-charge dispersion but, however, the superconducting phase is delocalized. Accordingly, the gate-charge dependence is smeared by the free rotations of the phase degree of freedom. 

We also note that the photon number response displays two sharp peaks for $n_{\rm g}=0.5$ at $A/\kappa \approx 13$ and $A/\kappa \approx 25$. The locations of the peaks are very sensitive to the value of the gate charge, i.e. to the energy level structure of the transmon. Similar abrupt changes in the transmon occupation were also observed in recent experiments in Ref.~\cite{lescanne2018}. They could be related to quantum chaotic motion of the system recently discussed in Ref.~\cite{mourik2018}.
In this parameter regime, also the Jaynes--Cummings model displays bistability~\cite{Vukics2018}.

\subsection{Comparison between different master equations}

\begin{figure}[ht!]
\includegraphics[width=1.0\linewidth]{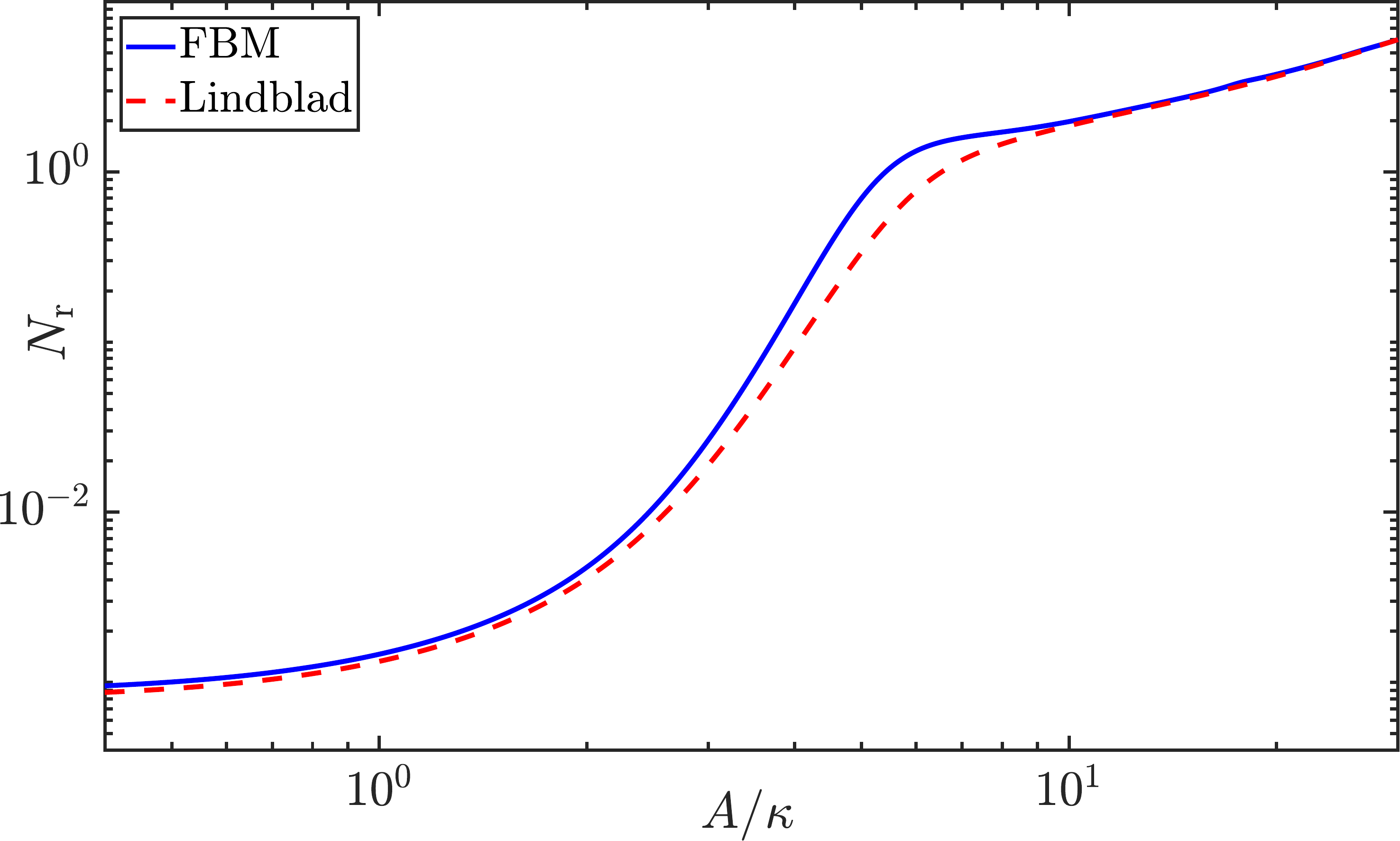}\hfill
\caption{Comparison between the Floquet--Born--Markov (FBM) and Lindblad models for dissipation in the two-level approximation for the transmon. The drive detuning is $\delta_{\rm d}/\omega_{\rm r}= -0.02$ and also the other 
parameters are the same as in Table~\ref{tab:params1}.}\label{fig:comparison}
\end{figure}

We have also compared our numerical Floquet--Born--Markov method against the Lindblad master equation which was presented in Sec.~\ref{sec:Lindblad} and has been conventionally used in the studies of similar strongly driven systems with weak dissipation. We note that for the case of strong coupling to the bath, the possible treatment is by the method of path integrals, as developed by Feynman and Vernon, which has already been applied to describe the dynamics of the Rabi model~\cite{Henriet2014}.

We recall that in the Lindblad formalism, the environment induces transitions between the non-driven states of the system, whereas in the Floquet--Born--Markov approach the dissipation couples to the drive-dressed states of the system. Thus, one expects deviations from the Floquet--Born--Markov results in the limit of strong driving. 
In Fig.~\ref{fig:comparison} we show a comparison between the two models in the two-state truncation approximation for the transmon. We see that the largest differences between the  models appear when the transition from the quantum to classical response starts to emerge, 
see Fig.~\ref{fig:Occupation7}. Based on our numerical calculations, the differences are the largest at resonance and both models give equivalent results 
whenever one of the three frequencies, $\omega_{\rm r}$, $\omega_{\rm q}$, or $\omega_{\rm d}$, is detuned from the other two.

We emphasize, however, that computationally the Floquet--Born--Markov master equation is by two orders of magnitude more efficient than the corresponding Lindblad equation. Moreover, in the case of Fig.~\ref{fig:DriveDetuning}(d) the computing time of the Floquet--Born--Markov equation was roughly a week with an ordinary CPU. 
In such cases, the solution of the Lindblad equation becomes impractical and one should use a parallelized implementation of 
the Floquet--Born--Markov master equation.

\section{Conclusions}\label{sec:V}

We have given a comprehensive treatment of the driven-dissipative quantum-to-classical phase transition for a Josephson pendulum coupled to a resonator, going beyond the truncated Rabi form of the Hamiltonian through the full inclusion of the higher energy level of the pendulum. \added{We refered to this as the multilevel Josephson--Rabi model.}
We modelled the open quantum system 
with the Floquet--Born--Markov method, in which 
the dissipative transitions occur between the drive-dressed states of the system. We compared our results also against those given by the conventional Lindblad formalism where the dissipation couples to the eigenstates of the non-driven system.

We found that the quantitative description of the multi-photon blockade phenomenon and of the nonlinearities associated with the phase transition in this system requires a systematic inclusion of the higher energy levels of the transmon and a proper model for dissipation. 
We also studied approximate classical models for this system, and showed that the discrete energy structure of the quantum system suppresses the classical chaotic motion of the quantum pendulum. Indeed, while the classical solution predicts a sudden change between the low and high amplitude solutions, the quantum solution displays a continuous transition from the normal-mode oscillations to the freely rotating pendulum regime. Finally, we analyzed in detail the two models of dissipation and demonstrated that they produce slightly different predictions for the onset of the photon blockade. 

\section{Acknowledgments} We thank D. Angelakis, S. Laine, and M. Silveri for useful discusssions. 
We would like to acknowledge financial support from the Academy of Finland under its Centre of Excellence program (projects 312296, 312057, 312298, 312300) and the Finnish Cultural Foundation. This work uses the facilities of the Low Temperature Laboratory (part of OtaNano) at Aalto University.

\appendix

\section{The eigenvalue problem for the Josephson pendulum}\label{app:eigenvalue}

The energy eigenstates of the pendulum can be solved from the Mathieu equation~\cite{baker2005,cottet2002,abramovitz1972} which produces a spectrum with bound and free-particle parts. The high-energy unbound states are given by the doubly-degenerate quantum rotor states, which are also the eigenstates of the angular momentum operator.

In analogy with the elimination of the vector potential by a gauge transformation as usually done for a particle in magnetic field, one can remove the dependence on $n_{\rm g}$ from the transmon Hamiltonian in Eq.~(\ref{eq:transmonHam}), i.e.
\begin{equation}
\hat H_{\rm t} = 4E_{\rm C}(\hat n-n_{\rm g})^2 - E_{\rm J}\cos \hat \varphi, 
 \end{equation}
with the gauge transformation $\hat U \hat H_{\rm t} \hat U^{\dag}$, where
\begin{equation}
\hat U = e^{-i n_{\rm g}\hat \varphi}.
\end{equation}
As a consequence, the eigenstates $|k\rangle$ of the Hamiltonian are modified into
\begin{equation}
|k\rangle \ \rightarrow \ e^{-in_{\rm g} \hat \varphi}|k\rangle.
\end{equation}
The transformed Hamiltonian can be written as
\begin{equation}
\hat H_{\rm t} = 4E_{\rm C}\hat n^2-E_{\rm J}\cos\hat \varphi.
\end{equation}
Here, we represent the  (Schr\"odinger) eigenvalue equation for the transformed Hamiltonian in the eigenbasis of the operator $\hat \varphi$.
As a result, the energy levels of the transmon can be obtained from the Mathieu equation~\cite{baker2005,cottet2002,abramovitz1972}
\begin{equation}\label{eq:Mathieu}
\frac{\partial^2}{\partial z^2}\psi_k(z) - 2 q \cos(2z) \psi_k(z) = -a\psi_k(z),
\end{equation}
where $z = \varphi/2$, $q=-\eta/2 = -E_{\rm J}/(2 E_{\rm C})$, and $a=E_k/E_{\rm C}$. 
We have also denoted the transformed eigenstate $|k\rangle$ in the $\varphi$ representation with $\psi_k(\varphi) = \langle \varphi | e^{-in_{\rm g} \hat \varphi}|k\rangle$. 
Note that $\Psi_k(\varphi) = e^{in_{\rm g}\varphi} \psi_k(\varphi)$ is the eigenfunction of the original Hamiltonian in Eq.~(\ref{eq:transmonHam}). 
Due to the periodic boundary conditions, one has that $\Psi_k(\varphi+2\pi) = \Psi_k(\varphi)$.
The solutions to Eq.~(\ref{eq:Mathieu}) are generally Mathieu functions which have a power series representation, but cannot be written in terms of elementary functions~\cite{abramovitz1972}. However, the corresponding energy-level structure can be studied analytically in the high and low-energy limits.

\begin{figure}[h!]
\includegraphics[width=\linewidth]{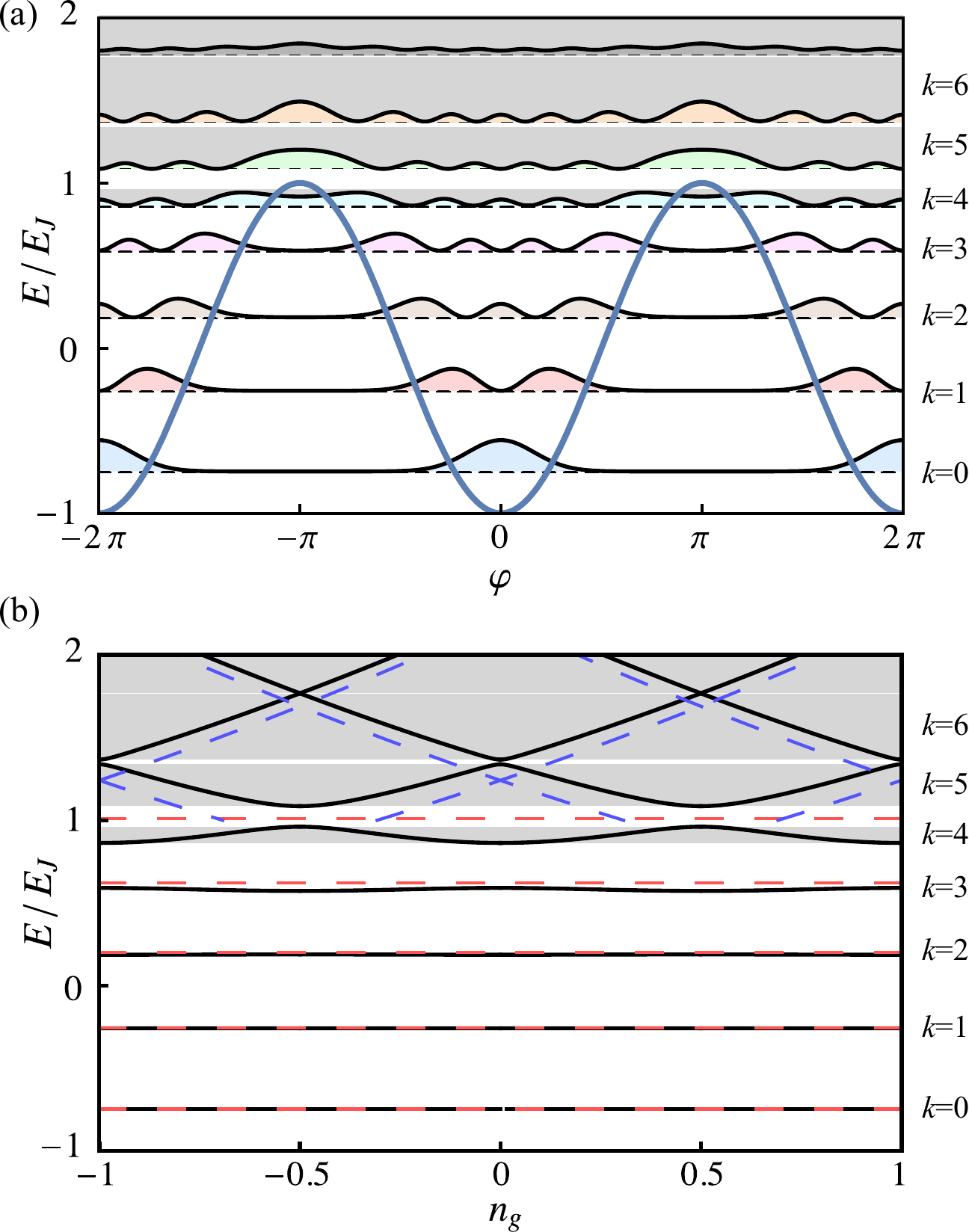}
\caption{
Eigenvalues and eigenstates of the transmon obtained with the Mathieu equation~(\ref{eq:Mathieu}) for $\eta =30$. 
(a) Eigenenergies as a function of the superconducting phase difference $\varphi$. The cosine potential is indicated with the blue \added{(gray)} line. Inside the well the eigenenergies are discrete and denoted with dashed black lines. On top of each line, we show the absolute square of the corresponding Mathieu eigenfunction. The energy bands from (b) are indicated with gray.		
(b) Eigenenergies as a function of the gate charge. We compare the numerically exact eigenenergies $E_k$ (solid black) with those of the perturbative Duffing oscillator (dashed \added{horizontal} red) and the free rotor (dashed \added{diagonal} blue). The charge dispersion in the (nearly) free rotor states leads to energy bands, which are denoted with gray. We show the Duffing and free rotor solutions only inside and outside the potential, respectively.  }\label{fig:Mathieuevals}
\end{figure}

In Fig.~\ref{fig:Mathieuevals}, we present the eigenenergies $E_k$ obtained as solutions of the Mathieu equation~(\ref{eq:Mathieu}). 
The eigenstates that lie within the wells formed by the cosine potential are localized in the coordinate $\varphi$, whereas the states far above are (nearly) evenly distributed, see Fig.~\ref{fig:Mathieuevals}(a). As a consequence, the high-energy states are localized in the charge basis.
The data shows that if plotted as a function of the gate charge, the states inside the cosine potential are nearly flat, see Fig.~\ref{fig:Mathieuevals}(b). This implies that such levels are immune to gate charge fluctuations, which results in a high coherence of the device. Outside the well, the energy dispersion with respect to the gate charge becomes significant, and leads to the formation of a band structure typical for periodic potentials~\cite{marder2000}. 

\subsection{High-energy limit: Free rotor}

If the energy in the system is very high due to, e.g., strong driving, the Josephson energy can be neglected and the transmon behaves as a free particle rotating in a planar circular orbit, which can be described solely by its angular momentum $\hat L_{\rm z} = \hat n$. 
Since the angular momentum is a good quantum number, the eigenenergies and the corresponding eigenfunctions are given by
\begin{equation}\label{eq:rotEn}
E_k = 4E_{\rm C}(k-n_{\rm g})^2, \ \ \psi_k(\varphi) = e^{i (k-n_{\rm g}) \varphi},
\end{equation}
where $k=0, \pm 1, \pm 2, \ldots$.
We note that if the magnetic field is zero ($n_g=0$), the nonzero free rotor energies are doubly degenerate. 
The level spacing is not constant but increases with increasing $k$ as~\cite{baker2005}
\begin{equation}
\Delta E_k = E_{k+1}-E_k = 4E_{\rm C}[ 2(k-n_{\rm g})+1].
\end{equation}

In Fig.~\ref{fig:Mathieuevals}, we show the eigenenergies calculated with Eq.~(\ref{eq:rotEn}). Clearly, with large energies outside the potential, the energy spectrum of the particle starts to resemble that of the free rotor.
Also, the eigenfunctions of the free rotor are plane waves in the $\varphi$ eigenbasis, yielding a flat probability density as a function of $\varphi$. 
On the other hand, in the momentum eigenbasis, the free rotor states are fully localized.

\subsection{Low-energy limit: Duffing oscillator}

If the pendulum energy is very low, the superconducting phase of the transmon is localized near $\varphi\approx 0$.
Thus, the cosine potential can be approximated with the first terms of its Taylor expansion.
Consequently, the transmon Hamiltonian reduces to that of a harmonic oscillator with an additional quartic potential
\begin{equation}
\hat H_{\rm t} \approx 4E_{\rm C}\hat n^2 + E_{\rm J}\left[-1 + \frac12\hat\varphi^2 - \frac{1}{12}\hat \varphi^4\right].
\end{equation}
This is the Hamiltonian operator of the quantum Duffing model. The Duffing model has received a considerable attention in the recent literature~\cite{Peano2006,Serban2007,Verso2010,Vierheilig2010,divincenzo2012, Everitt2005b} 
especially in the context of superconducting transmon realizations.
It is worthwhile to notice that in this regime the potential is no longer periodic and, thus, we can neglect the periodic boundary condition of the wavefunction. As a consequence, the eigenenergies and eigenfunctions are not dependent on the offset charge $n_{\rm g}$.

If $\eta = E_{J}/E_{C}\gg 1$, the quartic term is small and one can solve the eigenvalues and the corresponding eigenvectors perturbatively up to the first order in $\eta$. 
This regime in which 
the Josephson energy dominates over the charging energy is referred to as the transmon limit. 
One, therefore, obtains the eigenenergies
\begin{equation}\label{eq:DuffEn}
\frac{E_k}{4E_{\rm C}} = -\frac{\eta}{4}+\sqrt{\eta/2}\left(k+\frac12\right) - \frac{1}{48}(6k^2+6k+3),
\end{equation}
where $k=0,1,2,\ldots$.
Especially, the transition energy between the two lowest Duffing oscillator states can be written as
\begin{equation}\label{eq:DOqubitEn}
\hbar \omega_{\rm q} = E_1-E_0 = \sqrt{8E_{\rm J}E_{\rm C}}-E_{\rm C}.
\end{equation}
This becomes accurate as $\eta\rightarrow \infty$. The anharmonicity of a nonlinear oscillator is typically characterized in terms of the absolute and relative anharmonicity, which are defined, respectively, as
\begin{equation}
\mu = E_{12} - E_{01}\approx - E_{\rm C}, \ \ \mu_{\rm r} = \mu/E_{01}\approx -(8\eta)^{-1/2},
\end{equation}
where $E_{ij} = E_j-E_i$ and the latter approximations are valid in the transmon limit.
We emphasize that in the low-excitation limit the transmon oscillates coherently with frequency $\omega_{\rm q} \approx \omega_{\rm p} - E_{\rm C}/\hbar$. Thus, in the quantum pendulum the nonlinearity is present even in the zero-point energy, whereas the small-amplitude oscillations in classical pendulum occur at the angular (plasma) frequency $\hbar\omega_{\rm p} = \sqrt{8E_{\rm J}E_{\rm C}}$. 

In Fig.~(\ref{fig:Mathieuevals}), we compare the eigenenergies~(\ref{eq:DuffEn}) of the Duffing model obtained with the perturbation theory against the exact solutions of the Mathieu equation~(\ref{eq:Mathieu}). We see that in the low-energy subspace the perturbed-Duffing solution reproduces very well the full Mathieu results. For the higher excited states, the momentum dispersion starts to play a dominant role and deviations arise as expected. This starts to occur close to the boundary of the potential. One can estimate the number $K_{\rm b}$ of bound states by requiring $E_{K_{\rm b}-1}\approx E_{\rm J}$ in Eq.~(\ref{eq:DuffEn}).
This implies that the number of states within the potential scales with $\eta\gg 1$ as
\begin{equation}\label{eq:Nbound}
K_{\rm b} \propto \sqrt{\eta}.
\end{equation}
For the device with parameters listed in Table \ref{tab:params1}, one has that $\sqrt{\eta} \approx 5$, and the above estimate gives $K_{\rm b} \approx 5$. 
This coincides with the number of bound states extracted from the 
numerically exact spectrum of the eigenenergies depicted in Fig.~\ref{fig:Mathieuevals}. 

\section{Driven and damped classical system}\label{app:classeom}

The classical behaviour of the uncoupled pendulum has been extensively studied in the literature~\cite{Dykman1990, Dykman2012, baker2005}.
If the driving force is not too strong, one can approximate
the pendulum with a Duffing oscillator with a quartic non-linearity, as shown in the previous appendix. The main feature of such an oscillator is the bistability of its dynamics.  
Namely, in a certain range of drive amplitudes and frequency detunings between the driving signal
and the oscillator, two stable solutions with low and high amplitudes of the oscillations are possible. 
If one gradually increases the driving,
the pendulum suddenly jumps from the low to the high amplitude solution
at the critical driving strength, at which the low amplitude solution vanishes. 
In the bistable region the Duffing oscillator may
switch between the two solutions if one includes noise into the model \cite{Dykman1990}. This complicated dynamics has been observed in a classical Josephson junction~\cite{Siddiqi2004, Siddiqi2005}. 

However, differently from the previous work mentioned above, in our setup the pendulum is coupled to a resonator and driven only indirectly. Here we develop the classical theory of the coupled system and show that the basic physics of bistability is present as well. We first linearize the equations of motion and then introduce systematically the corrections due to the nonlinearity. The system Hamiltonian $\hat H_{\rm S} = \hat H_0 + \hat H_{\rm d}$, defined by Eq.~(\ref{eq:H0}) and Eq.~(\ref{eq:Hd}), can be written in terms of the circuit variables as
\begin{eqnarray}\label{eq:classHam}
\hat H_{\rm S} &=& \frac{\hat q^2}{2C_{\rm r}} + \frac{\hat \phi^2}{2L_{\rm r}} + 4E_{\rm C}\hat n^2-E_{\rm J}\cos \hat \varphi + \tilde g \hat n \hat q \nonumber\\
&& + \tilde A \cos(\omega_{\rm d}t) \hat q,
\end{eqnarray}
Above, we have denoted the capacitance and inductance of the $LC$ resonator with $C_{\rm r}$ and $L_{\rm r}$, respectively, the effective coupling with $\tilde g=\hbar g/q_{\rm zp}$, the effective drive with $\tilde A=\hbar A/q_{\rm zp}$, and the zero-point fluctuations with $\phi_{\rm zp} = \sqrt{\hbar/(2C_{\rm r}\omega_{\rm r})}$ and $q_{\rm zp}=\sqrt{C_{\rm r}\hbar \omega_{\rm r}/2}$. Also, the resonance frequency of the bare resonator is defined as $\omega_{\rm r}=1/\sqrt{L_{\rm r}C_{\rm r}}$.

The corresponding equations of motion for the expectation values of the dimensionless operators $\hat\phi_{\rm r}=\hat \phi/\phi_{\rm zp}$ and $\hat q_{\rm r}=\hat q/q_{\rm zp}$ can be written as
\begin{eqnarray}
\dot{\phi}_{\rm r} &=& \omega_{\rm r} q_{\rm r} + 2g n + 2 A\cos(\omega_{\rm d}t)-\frac{\kappa}{2}\phi_{\rm r},\label{eq:classeom1}\\
\dot{q}_{\rm r} &=& -\omega_{\rm r} \phi_{\rm r}-\frac{\kappa}{2}q_{\rm r},\label{eq:classeom2}\\
\dot{\varphi} &=& \frac{8E_{\rm C}}{\hbar}n+gq_{\rm r}-\frac{\gamma}{2}\varphi,\\
\dot{n} &=& -\frac{E_{\rm J}}{\hbar}\sin\varphi-\frac{\gamma}{2}n.\label{eq:classeom4}
\end{eqnarray}
where we have denoted the expectation value of operator $\hat x$ as $\langle \hat x\rangle\equiv x$, applied the commutation relations $[\hat \phi_{\rm r},\hat q_{\rm r}] = 2i$ and $[\hat\varphi,\hat{n}]=i$, and defined the phenomenological damping constants $\kappa$ and $\gamma = \gamma_0\omega_{\rm q}$ for the oscillator and the pendulum, respectively. The exact solution to these equations of motion is unavoidably numerical and is given in Figs.~\ref{fig:classsteps} and \ref{fig:Occupation7}. The resonator occupation is calculated as $N_{\rm r} = \frac{1}{4}(q_{\rm r}^2 +\phi_{\rm r}^2)$.

\subsection{Solution of the linearized equation}\label{app:lineom}

We study Eqs.~(\ref{eq:classeom1})-(\ref{eq:classeom4}) in the limit of weak driving. As a consequence, one can linearize the equations of motion by writing $\sin\varphi \approx \varphi$. In addition, by defining
\begin{eqnarray}
\alpha &=& \frac12(q_{\rm r}-i\phi_{\rm r}),\\
\beta &=& \frac{1}{\sqrt{2}}\left(\sqrt[4]{\frac{\eta}{8}} \varphi + i\sqrt[4]{\frac{8}{\eta}} n\right),
\end{eqnarray}
we obtain
\begin{equation}\begin{split}
\dot{\alpha}=& -i\omega_{\rm r}\alpha + g_{\rm eff}(\beta^*-\beta)-\frac{iA}{2}\left(e^{i\omega_{\rm d}t}+e^{-i\omega_{\rm d}t}\right) - \frac{\kappa}{2}\alpha,\\
\dot{\beta}=& - i\omega_{\rm p}\beta + g_{\rm eff}(\alpha+\alpha^*) - \frac{\gamma}{2}\beta,
\end{split}\end{equation}
where we have introduced an effective coupling as $g_{\rm eff}=g\sqrt[4]{\eta/32}$. The above equations describe two driven and dissipative coupled oscillators.

We assume that both oscillators are excited at the drive frequency, i.e. $\alpha = \alpha_0 \exp(-i\omega_{\rm d}t)$ and $\beta = \beta_0 \exp(-i\omega_{\rm d}t)$. By making a rotating-wave approximation for the coupling and the drive, we obtain the resonator occupation $N_{\rm lin} = |\alpha_0|^2$ in the steady state
\begin{equation}
N_{\rm lin} = \frac{A^2}{4}\frac{1}{\left(\delta_{\rm d}-g_{\rm eff}^2\frac{\delta_{\rm p}}{\delta_{\rm p}^2+\gamma^2/4}\right)^2+\left(\frac{\kappa}{2}+g_{\rm eff}^2\frac{\gamma/2}{\delta_{\rm p}^2+\gamma^2/4}\right)^2},
\end{equation}
where $\delta_{\rm d}=\omega_{\rm d}-\omega_{\rm r}$ and $\delta_{\rm p} = \omega_{\rm d}-\omega_{\rm p}$, with $\omega_{\rm p}=\sqrt{8E_{\rm J}E_{\rm C}}/\hbar$. This appeared already in Eq.~(\ref{eq:linearNr}).

\subsection{Correction due to the pendulum nonlinearity}\label{app:nonlin}

Here, we study the nonlinear effects neglected in the above linearized calculation. We eliminate the variables $\phi_{\rm r}$ and $n$ from Eqs.~(\ref{eq:classeom1})-(\ref{eq:classeom4}) and obtain
\begin{eqnarray}
\ddot{q}_{\rm r} + \kappa \dot{q}_{\rm r} + \tilde{\omega}_{\rm r}^2q_{\rm r}+g_1 \dot{\varphi} + 2A\omega_{\rm r}\cos(\omega_{\rm d}t) &=& 0,\label{eq:classqr}\\
\ddot{\varphi} + \gamma\dot{\varphi}+\omega_{\rm p}^2\sin\varphi  - g\dot{q}_{\rm r}&=& 0,\label{eq:classvarphi}
\end{eqnarray}
where we have denoted $g_1=g \hbar \omega_{\rm r}/(4E_{\rm C})$, 
and defined the renormalized resonator frequency as $\tilde{\omega}_{\rm r}^2 = \omega_{\rm r}^2 - g^2\hbar \omega_{\rm r}/(4E_{\rm C})$. In Eq.~(\ref{eq:classqr}), we have included only the term that is proportional to $g^2$ as it provides the major contribution to the frequency renormalization, and neglected the other second order terms in $\kappa$, $\gamma$ and $g$ that lead to similar but considerable smaller effects.
We write the solutions formally in terms of Fourier transform as
\begin{eqnarray}
q_{\rm r}(t) &=& \int \frac{d\Omega}{2\pi}q_{\rm r}[\Omega]e^{-i\Omega t},\\
\varphi(t) &=& \int \frac{d\Omega}{2\pi}\varphi[\Omega]e^{-i\Omega t},
\end{eqnarray}
where $q_{\rm r}[\Omega]$ and $\varphi[\Omega]$ are the (complex valued) Fourier coefficients of $q_{\rm r}(t)$ and $\varphi(t)$, respectively. 
As a consequence, one can write the equations of motion as
\begin{widetext}
\begin{eqnarray}
\int \frac{d\Omega}{2\pi}\left\{\left(\tilde{\omega}_{\rm r}^2 - \Omega^2 - i\kappa\Omega\right)q_{\rm r}[\Omega] - ig_1\Omega \varphi[\Omega]+2\pi A\omega_{\rm r} \left[\delta(\Omega-\omega_{\rm d})+\delta(\Omega+\omega_{\rm d})\right]\right\}e^{-i\Omega t}&=&0,\\
\int \frac{d\Omega}{2\pi}\left\{\left(- \Omega^2 - i\gamma\Omega\right)\varphi[\Omega] + ig\Omega q_{\rm r}[\Omega]\right\}e^{-i\Omega t} + \omega_{\rm p}^2\sin\varphi &=&0.\label{eq:classvphi}
\end{eqnarray}
We solve $q_{\rm r}[\Omega]$ from the first equation and obtain
\begin{eqnarray}\label{eq:classqr2}
q_{\rm r}[\Omega] &=& \frac{ig_1\Omega\varphi[\Omega]-2\pi A\omega_{\rm r}[\delta(\Omega-\omega_{\rm d})+\delta(\Omega+\omega_{\rm d})]}{\tilde{\omega}_{\rm r}^2-\Omega^2-i\kappa\Omega}.
\end{eqnarray}
By replacing this result into Eq.~(\ref{eq:classvphi}), we obtain
\begin{equation}
\int \frac{d\Omega}{2\pi}\left\{\left(- \Omega^2 - i\gamma\Omega- \frac{gg_1\Omega^2}{\tilde{\omega}_{\rm r}^2-\Omega^2-i\kappa\Omega}\right)\varphi[\Omega] \right\}e^{-i\Omega t} + \omega_{\rm p}^2\sin\varphi =\frac{2gA\omega_{\rm r}\omega_{\rm d}}{\sqrt{(\tilde{\omega}_{\rm r}^2-\omega_{\rm d}^2)^2+\kappa^2\omega_{\rm d}^2}}\cos(\omega_{\rm d}t),
\end{equation}
\end{widetext}
where we have neglected a constant phase factor. 
For weak drive amplitudes, $\varphi[\omega_{\rm d}]$ is the only non-zero Fourier component. Thus, one can evaluate the Fourier transform in the above equation at the drive frequency. Consequently, the Fourier component of the third term in the equation can be evaluated as
\begin{equation}\begin{split}
\frac{gg_1\Omega^2}{\tilde{\omega}_{\rm r}^2-\Omega^2-i\kappa\Omega} &\approx \frac{gg_1\omega_{\rm d}^2}{(\tilde{\omega}_{\rm r}^2-\omega_{\rm d}^2)^2+\kappa^2\omega_{\rm d}^2}\left[(\tilde{\omega}_{\rm r}^2-\omega_{\rm d}^2)+i\kappa\omega_{\rm d}\right]\\
&\approx \frac{gg_1\omega_{\rm d}^2}{\tilde{\omega}_{\rm r}^2-\omega_{\rm d}^2} + i\frac{gg_1\kappa\omega_{\rm d}^3}{(\tilde{\omega}_{\rm r}^2-\omega_{\rm d}^2)^2+\kappa^2\omega_{\rm d}^2},
\end{split}\end{equation}
where in the second term we have assumed that the dissipation is weak, i.e. $\kappa\ll\sqrt{\tilde\omega_{\rm r}^2-\omega_{\rm d}^2}$ and we taken into account the dominant terms for the real and imaginary parts. As a result, we obtain
\begin{equation}
\ddot{\varphi}+\tilde{\gamma}\dot{\varphi}+\omega_{\rm p}^2\sin\varphi + (\tilde{\omega}_{\rm p}^2-\omega_{\rm p}^2)\varphi = B\cos(\omega_{\rm d} t). \label{eqn:appendix_nonlin}
\end{equation}
Here, we have defined the renormalized linear oscillation frequency $\tilde{\omega}_{\rm p}$, dissipation rate $\tilde{\gamma}$, and drive amplitude $B$ as
\begin{eqnarray}
\tilde{\omega}_{\rm p}^2 &=& \omega_{\rm p}^2-\frac{g g_1 \omega_{\rm d}^2}{\tilde{\omega}_{\rm r}^2-\omega_{\rm d}^2},\\
\tilde{\gamma} &=& \gamma+\frac{gg_1\kappa\omega_{\rm d}^2}{(\tilde{\omega}_{\rm r}^2-\omega_{\rm d}^2)^2+\kappa^2\omega_{\rm d}^2},\\
B &=& \frac{2gA\omega_{\rm r}\omega_{\rm d}}{\sqrt{(\tilde{\omega}_{\rm r}^2-\omega_{\rm d}^2)^2+\kappa^2\omega_{\rm d}^2}},\label{eq:effamp}
\end{eqnarray}
where the two first equations are valid if $\kappa \ll \sqrt{\tilde\omega_{\rm r}^2-\omega_{\rm d}^2}$.

Thus, we have shown that in the limit of low dissipation, the classical resonator-transmon system can be modeled as a driven and damped pendulum. In the case of weak driving, we expand the sinusoidal term up to the third order in $\varphi$. We obtain the equation of motion for the driven and damped Duffing oscillator:
\begin{equation}\label{eq:duff}
\ddot{\varphi}+\tilde{\gamma}\dot{\varphi}+\tilde{\omega}_{\rm p}^2\left(\varphi-\frac{\omega_{\rm p}^2}{6\tilde{\omega}_{\rm p}^2}\varphi^3\right) = B\cos(\omega_{\rm d} t).
\end{equation}
This equation can be solved approximatively by applying a trial solution $\varphi(t) = \varphi_1 \cos(\omega_{\rm d}t)$ into Eq.~(\ref{eq:duff}). By applying harmonic balance, and neglecting 
super-harmonic terms, we obtain a relation for the amplitude $\varphi_1$ in terms of the drive amplitude $B$. 
By taking a second power of this equation and, again, neglecting the super-harmonic terms, we obtain
\begin{equation}
\left[\left(\tilde{\omega}_{\rm p}^2-\omega_{\rm d}^2-\frac{\omega_{\rm p}^2}{8}\varphi_1^2\right)^2+\tilde\gamma^2\omega_{\rm d}^2\right]\varphi_1^2 = B^2.
\end{equation}
The above equation is cubic in $\varphi_1^2$. It has one real solution if the discriminant $D$ of the equation is negative, i.e. $D<0$. If $D>0$, the equation has three real solutions, 
two stable and one unstable. The stable solutions can appear only if $\omega_{\rm d}<\tilde{\omega}_{\rm p}$, which is typical for Duffing oscillators with a soft spring (negative nonlinearity). The bistability can, thus, occur for amplitudes $B_{\rm min}<B<B_{\rm crit}$ where the minimal and critical amplitudes $B_{\rm min}$ and $B_{\rm crit}$, respectively, determine the region of bistability and are obtained from the equation $D=0$. By expanding the resulting $B_{\rm min}$ and $B_{\rm crit}$ in terms of $\tilde\gamma$ and by taking into account the dominant terms, we find that
\begin{eqnarray}
B_{\rm min} &=& \tilde\gamma\frac{\omega_{\rm d}}{\omega_{\rm p}}\sqrt{8(\tilde\omega_{\rm p}^2-\omega_{\rm d}^2)} = \tilde\gamma\frac{\sqrt{27}\omega_{\rm d}}{2(\tilde\omega_{\rm p}^2-\omega_{\rm d}^2)}B_{\rm crit}\label{eq:bistabmin}\\
B_{\rm crit} &=& \sqrt{\frac{32}{27}}\frac{(\tilde{\omega}_{\rm p}^2-\omega_{\rm d}^2)^{3/2}}{\omega_{\rm p}}\approx \frac{16}{3\sqrt{3}}\sqrt{\omega_{\rm p}\delta_{\rm p}^3},\label{eq:anjump}
\end{eqnarray}
where the last equality holds if $\delta_{\rm p} = \tilde{\omega}_{\rm p}-\omega_{\rm d}\ll \omega_{\rm p}$. 
The iterative numerical solution of Eq.~(\ref{eq:duff}) indicates that the initial state affects the switching location between the two stable solutions. 
We note that this approximation neglects all higher harmonics and, thus, cannot reproduce any traces towards chaotic motion inherent to the strongly driven pendulum. 

Finally, we are able to write the minimal and critical drive amplitudes of the coupled resonator-transmon system using Eqs.~(\ref{eq:effamp}),~(\ref{eq:bistabmin}), and~(\ref{eq:anjump}). We obtain [see Eq.~(\ref{eq:duffan})]
\begin{eqnarray}
A_{\rm min} &=& \tilde\gamma\sqrt{2(\tilde\omega_{\rm p}^2-\omega_{\rm d}^2)}\frac{\sqrt{(\tilde{\omega}_{\rm r}^2-\omega_{\rm d}^2)^2+\kappa^2\omega_{\rm d}^2}}{g\omega_{\rm r}\omega_{\rm p}},\\
A_{\rm crit} &=& \sqrt{\frac{8}{27}}	\left(\tilde{\omega}_{\rm p}^2-\omega_{\rm d}^2\right)^{3/2}\frac{\sqrt{(\tilde{\omega}_{\rm r}^2-\omega_{\rm d}^2)^2+\kappa^2\omega_{\rm d}^2}}{g\omega_{\rm r}\omega_{\rm d}\omega_{\rm p}}.
\end{eqnarray}
Note that these equations are valid for $\kappa \ll \sqrt{\tilde\omega_{\rm q}^2-\omega_{\rm d}^2}$.


\bibliography{nonlinear_bib}{}

\end{document}